# The variance of identity-by-descent sharing in the Wright-Fisher model

Shai Carmi,[1] Pier Francesco Palamara,[1] Vladimir Vacic,[1] Todd Lencz,[2, 3] Ariel Darvasi,[4] and Itsik Pe'er[5, 6]

[1]*Department of Computer Science, Columbia University, New York, NY 10027, USA*
[2]*Department of Psychiatry, Division of Research,*
*The Zucker Hillside Division of the North Shore-Long Island Jewish Health System, Glen Oaks, NY, 11004, USA*
[3]*Center for Psychiatric Neuroscience, The Feinstein Institute for Medical Research, NSLIJ Health System, Manhasset, NY*
[4]*Department of Genetics, The Institute of Life Sciences,*
*The Hebrew University of Jerusalem, Givat Ram, Jerusalem, 91904, Israel*
[5]*Department of Computer Science, Columbia University, New York, NY 10025, USA*
[6]*Center for Computational Biology and Bioinformatics,*
*Columbia University, New York, NY, 10032, USA*

(Dated: August 12, 2013)

Widespread sharing of long, identical-by-descent (IBD) genetic segments is a hallmark of populations that have experienced recent genetic drift. Detection of these IBD segments has recently become feasible, enabling a wide range of applications from phasing and imputation to demographic inference. Here, we study the distribution of IBD sharing in the Wright-Fisher model. Specifically, using coalescent theory, we calculate the variance of the total sharing between random pairs of individuals. We then investigate the cohort-averaged sharing: the average total sharing between one individual and the rest of the cohort. We find that for large cohorts, the cohort-averaged sharing is distributed approximately normally. Surprisingly, the variance of this distribution does not vanish even for large cohorts, implying the existence of "hyper-sharing" individuals. The presence of such individuals has consequences for the design of sequencing studies, since, if they are selected for whole-genome sequencing, a larger fraction of the cohort can be subsequently imputed. We calculate the expected gain in power of imputation by IBD, and subsequently, in power to detect an association, when individuals are either randomly selected or specifically chosen to be the hyper-sharing individuals. Using our framework, we also compute the variance of an estimator of the population size that is based on the mean IBD sharing and the variance in the sharing between inbred siblings. Finally, we study IBD sharing in an admixture pulse model, and show that in the Ashkenazi Jewish population the admixture fraction is correlated with the cohort-averaged sharing.

## I. INTRODUCTION

In isolated populations, even purported unrelated individuals often share genetic material that is *identical-by-descent*, or IBD. Traditionally, the term IBD sharing referred to co-ancestry at a single site (or autozygosity, in the case of a diploid individual) and was widely investigated as a measure of the degree of inbreeding in a population [1]. Recent years have brought dramatic increases in the quantity and density of available genetic data, and together with new computational tools, enabled the detection of IBD sharing of entire genomic segments (see, e.g., [2–8]). The availability of IBD detection tools that are efficient enough to detect shared segments in large cohorts has resulted in numerous applications, from demographic inference [9, 10] to characterization of populations [11] to selection detection [12], relatedness detection and pedigree reconstruction [13–16], prioritization of individuals for sequencing [17], inference of HLA type [18], detection of haplotypes associated with a disease or a trait [19–21], imputation [22], and phasing [23].

Recently, some of us used coalescent theory to calculate several theoretical quantities of IBD sharing under a number of demographic histories. Then, shared segments were detected in real populations, and their demographic histories were inferred [10]. Here, we expand upon [10] to investigate additional aspects of the stochastic variation in IBD sharing. Specifically, we provide a precise calculation for the variance of the total sharing in the Wright-Fisher model, either between a random pair of individuals or between one individual and all others in the cohort.

Understanding the variation in IBD sharing is an important theoretical characterization of the Wright-Fisher model, and additionally, it has several practical applications. For example, it can be used to calculate the variance of an estimator of the population size that is based on the sharing between random pairs. In a different domain, the variance in IBD sharing is needed to accurately assess strategies for sequencing studies, specifically, in prioritization of individuals to be sequenced. This is because imputation strategies use IBD sharing between sequenced individuals and genotyped, not-sequenced individuals to increase the number of effective sequences analyzed in the association study [17, 22, 23].

In the remainder of the paper, we first review the derivation of the mean fraction of the genome shared between two individuals [10]. We then calculate the variance of this quantity using coalescent theory with recombination. We provide a number of approximations, one of which results in a surprisingly simple expression, which is then generalized to a variable population size and to the sharing of segments in a length range. We also numerically investigate the pairwise sharing distribution and provide an approximate fit. We then turn to the average total sharing between each individual and the entire cohort. We show that this quantity, which we term the cohort-averaged sharing, is approximately normally distributed, but is much wider than naively expected, implying the existence of hyper-sharing individuals. We consider several applications: the number of individuals needed to be sequenced to achieve a certain imputation power and the implications to disease mapping, inference of the population size based on the total sharing, and the variance of the sharing between siblings. We finally calculate the mean and the variance of the sharing in an admixture pulse model and show numerically that admixture results in a broader than expected cohort-averaged sharing. Therefore, large variance of the cohort-averaged sharing can indicate admixture. In the Ashkenazi Jewish population, we show that the cohort-averaged sharing is strongly anti-correlated with the fraction of European ancestry.

## II. RESULTS

### A. Variation in IBD sharing in the Wright-Fisher model

#### 1. Definitions

*The Wright-Fisher model.* — We consider the standard Wright-Fisher model for a finite, isolated population, described by $2N$ haploid chromosomes, where each pair of chromosomes corresponds to one diploid individual. Each chromosome in the current generation descends, with equal probability, from one of the chromosomes in the previous generation, and recombination occurs at rate 0.01/cM per generation. The Wright-Fisher model has been widely investigated both in forward dynamics and under the coalescent [24]. For simplicity of notation, we denote the number of individuals, or the population size, as $N$, even though we really refer to the number of haploids and not the number of individuals. Throughout most of the analysis, we assume that each individual carries a single chromosome of length $L$ centiMorgans (cM).

*IBD sharing.* — We say that a genomic segment is shared, or is IBD, between two individuals if it is longer than $m$(cM) and it has been inherited without recombination from a single common ancestor. We do not require the shared segments to be completely identical. That is, if any mutation has occurred since the time of the most recent common ancestor (MRCA), that would not disqualify the segments from being shared IBD according to our definition. The reason is that even in the presence of mutations, an order of magnitude calculation shows that regardless of the segment length, two individuals sharing a segment are expected to differ in just $\approx 1$ site along the segment (see the Supplementary Material, Section S1.1). Therefore, in a long IBD segment, the number of differences should be very small compared to the number of matches. In practice, there are also other sources of error in IBD detection, most notably phase switch errors. We assume, however, that there always exists a large enough length threshold above which segments are detectable without errors [3, 7], which corresponds to the parameter $m$ introduced above; the precise value of the threshold will depend on the genotyping/sequencing technology. We assume that information is available for $M$ markers, uniformly distributed (in genetic distance) along the chromosome, and densely enough that any effect caused by the discreteness of the markers is negligible (say, if $m \cdot (M/L) \gg 1$). We define the *total sharing* between two individuals as the fraction of their markers that are found in shared segments.

#### 2. Mean total sharing

In this subsection, we review the derivation of the mean fraction of the genome found in segments shared between two individuals [10]. We assume that the coalescent process along the chromosome can be approximated by the Sequentially Markov Coalescent [25], and ignore the different behavior of sites at the ends of the chromosome. Consider first a single site $s$ and assume that its MRCA dates $g$ generations ago. The total length $\ell$ of the segment in which the site is found is the sum of $\ell_R$ and $\ell_L$, where $\ell_R$ and $\ell_L$ are the segment lengths to the right and left of $s$, respectively (all lengths are in cM). The distributions of $\ell_R$ and $\ell_L$ are exponential with rate $g/50$, since the two individuals were separated by $2g$ meioses, each of which introduces a recombination event with rate 0.01/cM, and the nearest recombination would terminate the shared segment. The probability $\pi$ of the total segment length, $\ell$, to exceed $m$ is, given $g$,

$$\pi|g = \int_m^\infty \ell \left(\frac{g}{50}\right)^2 e^{-\frac{g}{50}\ell} d\ell = \left(1 + \frac{mg}{50}\right) e^{-\frac{mg}{50}}. \quad (1)$$

According to coalescent theory in the Wright-Fisher model, under the continuous-time scaling $g \to Nt$ the times to the MRCA are exponentially distributed with rate 1: $\Phi(t) = e^{-t}$. Therefore,

$$\pi = \int_0^\infty e^{-t} \left(1 + \frac{mNt}{50}\right) e^{-\frac{mNt}{50}} dt$$
$$= \frac{100(25 + mN)}{(50 + mN)^2}. \quad (2)$$

The total fraction of the genome found in shared segments is

$$f_T = \frac{1}{M} \sum_{s=1}^{M} I(s), \quad (3)$$

where $I(s)$ is the indicator that site $s$ is in a shared segment, and the sum is over all sites. The mean fraction of the genome shared is

$$\langle f_T \rangle = \frac{1}{M} \sum_{s=1}^{M} \langle I(s) \rangle = \pi = \frac{100(25 + mN)}{(50 + mN)^2}, \quad (4)$$

where $\langle \cdot \rangle$ denotes the average over all ancestral processes. As expected, for $mN \to \infty$, $\langle f_T \rangle \to 0$ and for $mN \to 0$, $\langle f_T \rangle \to 1$. For large $N$, we have $\langle f_T \rangle \approx 100/(mN)$.

### 3. The variance of the total sharing

We now turn to calculating the variance of the total sharing. Using Eq. (3),

$$
\begin{aligned}
\mathrm{Var}\,[f_T] &= \mathrm{Var}\left[\frac{1}{M}\sum_{s=1}^{M}I(s)\right] \\
&= \frac{\pi(1-\pi)}{M} + \frac{1}{M^2}\sum_{s_1}\sum_{s_2 \neq s_1}\mathrm{Cov}\,[I(s_1), I(s_2)] \\
&= \frac{\pi(1-\pi)}{M} + \frac{1}{M^2}\sum_{s_1}\sum_{s_2 \neq s_1}\left[\pi_2(s_1, s_2) - \pi^2\right],
\end{aligned}
$$

where $\pi_2(s_1, s_2)$ is the probability that both markers $s_1$ and $s_2$ are on shared segments and $\pi$ is given by Eq. (2). In the rest of the section, we assume that each individual carries one chromosome only, for if we have $c$ chromosomes, each of length $L_i$, then (if the two individuals are not close relatives)

$$
f_T = \frac{\sum_{i=1}^{c} L_i f_{T,(i)}}{\sum_{i=1}^{c} L_i},
$$

and the variance is

$$
\mathrm{Var}\,[f_T] = \frac{\sum_{i=1}^{c} L_i^2 \mathrm{Var}\,[f_{T,(i)}]}{\left(\sum_{i=1}^{c} L_i\right)^2}, \tag{5}
$$

where $f_{T,(i)}$ is the total sharing in chromosome $i$, assumed independent of the other chromosomes. Rewriting $\pi_2(s_1, s_2)$ as $\pi_2(k)$, where $k$ is the number of markers separating $s_1$ and $s_2$, we have

$$
\mathrm{Var}\,[f_T] = \frac{\pi(1-\pi)}{M} + \frac{2}{M^2}\sum_{k=1}^{M}(M-k)\left[\pi_2(k) - \pi^2\right]. \tag{6}
$$

All is left to evaluate $\pi_2(k)$, for which we provide three approximations. The first is presented below and the second (which is a variation of the first) is presented in Supplementary Section S1.3. The third approximation, which is the most crude but which yields an explicit dependence on the population parameters, is presented in Section II A 4.

In the first approach, we assume that once the times $t_1, t_2$ to the MRCA at the two sites are known, the sites are (or are not) in shared segments independently of each other, and with probabilities given by Eq. (1). Clearly, this assumption is violated when both sites belong to the same shared segment, and in Supplementary Section S1.3, we show how this assumption can be avoided (but in the cost of significantly complicating the analysis). Nevertheless, it gives a good approximation, as we will later

see (Figure 2). We can therefore use Eq. (1) to write

$$
\begin{aligned}
\pi_2(k) &\approx \int_0^\infty \int_0^\infty dt_1 dt_2 \Phi(t_1, t_2) \times \\
&\quad \times \left(1 + \frac{mNt_1}{50}\right)e^{-\frac{mNt_1}{50}}\left(1 + \frac{mNt_2}{50}\right)e^{-\frac{mNt_2}{50}} \\
&= \widehat{\Phi}\left(\frac{mN}{50}, \frac{mN}{50}\right) - m\frac{\partial}{\partial m}\widehat{\Phi}\left(\frac{mN}{50}, \frac{mN}{50}\right) \\
&\quad + m^2\left[\frac{\partial}{\partial m_1}\frac{\partial}{\partial m_2}\widehat{\Phi}\left(\frac{m_1 N}{50}, \frac{m_2 N}{50}\right)\right]_{\substack{m_1 = m \\ m_2 = m}},
\end{aligned} \tag{7}
$$

where $\Phi(t_1, t_2)$ is the joint PDF (probability density function) of $t_1$ and $t_2$ and

$$
\widehat{\Phi}(q_1, q_2) = \int_0^\infty \int_0^\infty e^{-q_1 t_1 - q_2 t_2}\Phi(t_1, t_2)dt_1 dt_2
$$

is the Laplace transform of $\Phi(t_1, t_2)$. We therefore reduced the problem of finding $\pi_2(k)$ into that of finding $\widehat{\Phi}(q_1, q_2)$.

To find $\Phi(t_1, t_2)$ (or rather, its Laplace transform), we use the continuous-time Markov chain representation of the coalescent with recombination [24, 26, 27]. The chain is illustrated in Figure 1. Initially (present time), the chain is in state 1, corresponding to two chromosomes carrying two sites each. The chain terminates at state 8, when both sites have reached their MRCA. To construct the chain, coalescence events were assumed to occur at rate 1 and recombination events at rate $\rho/2$, where $\rho = 2N\frac{k}{M}L/100$ is the scaled recombination rate [24].

Denote by $P_i(t)$ the probability that the chain is at state $i$ at time $t$, given that it started at state 1. The probability that the two sites have reached their MRCA simultaneously in the time range $[t, t+dt]$ is $P_1(t)dt$, since this is the product of the probability that the chain is at state 1 at time $t$ ($P_1(t)$) and the probability of the transition $1 \to 8$ in the given time interval ($dt$). The probability that only the left site has reached its MRCA (and the right site has not) in $[t, t+dt]$ is $P_2(t)dt + P_3(t)dt$: this corresponds to the transitions $2 \to 5$ and $3 \to 7$. This is also the probability that only the right site has reached its MRCA in $[t, t+dt]$ (transitions $2 \to 4$, $3 \to 6$). Finally, the probability that the left site has reached its MRCA in $[t_1, t_1+dt_1]$ and that the right site has reached its MRCA in $[t_2, t_2+dt_2]$ ($t_2 > t_1$) is $[P_2(t_1) + P_3(t_1)]dt_1 e^{-(t_2-t_1)}dt_2$. This is true, because the exit rate from state 5 and 7 is 1; therefore, the probability that the chain will wait at one of those states for time $(t_2 - t_1)$ and then leave to the terminal state is $e^{-(t_2-t_1)}dt_2$. Similar considerations apply for the case $t_1 > t_2$ (with the transitions $2 \to 4$ and $3 \to 6$). In sum, for $t_1, t_2 > 0$,

$$
\begin{aligned}
\Phi(t_1, t_2) &= P_1(t_1)\delta(t_2 - t_1) \\
&\quad + [P_2(t_1) + P_3(t_1)]e^{-(t_2-t_1)}\Theta(t_2 - t_1) \\
&\quad + [P_2(t_2) + P_3(t_2)]e^{-(t_1-t_2)}\Theta(t_1 - t_2),
\end{aligned}
$$

where $\delta(t)$ is the Dirac delta function and $\Theta(t) = 1$ for $t > 0$ and is otherwise zero. Laplace transforming the last equation,

$$
\begin{aligned}
\widehat{\Phi}(q_1, q_2) &= \int_0^\infty \int_0^\infty e^{-q_1 t_1 - q_2 t_2} P_1(t_1) \delta(t_2 - t_1) dt_1 dt_2 \\
&+ \int_0^\infty \int_{t_1}^\infty e^{-q_1 t_1 - q_2 t_2} [P_2(t_1) + P_3(t_1)] e^{-(t_2 - t_1)} dt_2 dt_1 \\
&+ \int_0^\infty \int_{t_2}^\infty e^{-q_1 t_1 - q_2 t_2} [P_2(t_2) + P_3(t_2)] e^{-(t_1 - t_2)} dt_1 dt_2 \\
&= \widehat{P_1}(q_1 + q_2) \\
&+ \frac{1}{q_2 + 1} \int_0^\infty e^{-(q_1 + q_2) t_1} [P_2(t_1) + P_3(t_1)] dt_1 \\
&+ \frac{1}{q_1 + 1} \int_0^\infty e^{-(q_1 + q_2) t_2} [P_2(t_2) + P_3(t_2)] dt_2 \\
&= \widehat{P_1}(q_1 + q_2) \\
&+ \left[ \widehat{P_2}(q_1 + q_2) + \widehat{P_3}(q_1 + q_2) \right] \left[ \frac{1}{q_1 + 1} + \frac{1}{q_2 + 1} \right]. \quad (8)
\end{aligned}
$$

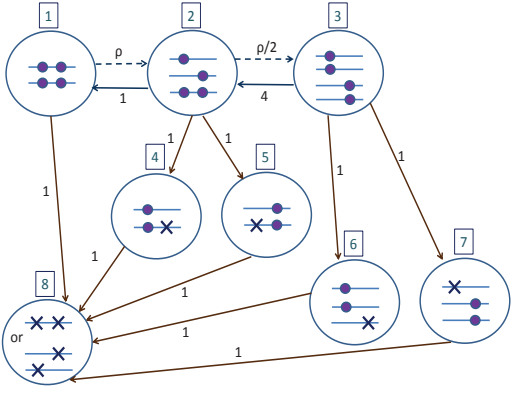

FIG. 1: **An illustration of the continuous-time Markov chain representation of the coalescent with recombination [24, 27].** Circles correspond to states, with the state number in a box on top of each circle. Arrows connecting circles represent transitions (solid lines: coalescence events; dashed lines: recombination events), with their rates indicated. The lines inside each circle represent chromosomes with two sites each. Ancestral sites are indicated as either small circles (as long as there are still two lineages carrying the ancestral material) or crosses (whenever the two lineages coalesced and the site has reached its MRCA). Transitions leading to the MRCA in one or two sites are colored brown. Transitions between states 4 and 6 and between 5 and 7 are not indicated, as they do not affect the final coalescence times. The schematic was adapted from [24].

In the last equation, $\widehat{P_i}(q) = \int_0^\infty e^{-qt} P_i(t) dt$ ($i = 1, 2, 3$) are the Laplace transforms of $P_i(t)$. The Laplace transforms can be calculated using the general relation [28]

$$
\widehat{P_i}(q) = (qI - Q)^{-1}_{1i}, \quad (9)
$$

where $Q$ is the transition rate matrix: $Q_{ij}$ is the transi-tion rate from $i$ to $j \neq i$ and $Q_{ii} = -\sum_{j \neq i} Q_{ij}$,

$$
Q = \begin{pmatrix}
-1 - \rho & \rho & 0 & 0 & 0 & 0 & 0 & 1 \\
1 & -3 - \rho/2 & \rho/2 & 1 & 1 & 0 & 0 & 0 \\
0 & 4 & -6 & 0 & 0 & 1 & 1 & 0 \\
0 & 0 & 0 & -1 & 0 & 0 & 0 & 1 \\
0 & 0 & 0 & 0 & -1 & 0 & 0 & 1 \\
0 & 0 & 0 & 0 & 0 & -1 & 0 & 1 \\
0 & 0 & 0 & 0 & 0 & 0 & -1 & 1 \\
0 & 0 & 0 & 0 & 0 & 0 & 0 & 0
\end{pmatrix}.
\quad (10)
$$

Using Eqs. (8), (9), and (10), and Mathematica,

$$
\widehat{\Phi}(q_1, q_2) = \frac{2AB + C(D + q_1 q_2) + E}{A[2(A - q_1 q_2)B + CD + E]}, \quad (11)
$$

where $A = (1 + q_1)(1 + q_2)$, $B = (3 + q_1 + q_2)(6 + q_1 + q_2)$, $C = \rho(2 + q_1 + q_2)$, $D = 13 + 3(q_1 + q_2)$ and $E = \rho^2(2 + q_1 + q_2)$. Eq. (11) was also derived in [29], using the birth-and-death-process equivalent of the coalescent with recombination, and can also be derived using the Feynman-Kac formula (see Supplementary Section S1.4). Substituting, using Mathematica, Eq. (11) in (7) and then using Eq. (6) gives an expression for the variance,

$$
\text{Var}[f_T] = \mathcal{F}(N, m, L, M). \quad (12)
$$

The function $\mathcal{F}$ is too long to reproduce here, but can be found in the Supplementary Matlab code (File S2). For further discussion on the approximations made, see Supplementary Section S1.2. The *standard deviation* (SD) of the total sharing is defined as usual as $\sigma_{f_T} \equiv \sqrt{\text{Var}[f_T]}$.

To evaluate the accuracy of our expressions for the mean and SD of the total sharing, we used the Genome coalescent simulator [30], along with an add-on that returns, for each generated genealogy, the locations of the segments that are IBD between each pair of individuals [10]. The simulation results (see also Methods) are presented and compared to the theory in Figure 2. In each panel, we varied one of $N$, $m$, and $L$, keeping the two others fixed (as long as the marker density is large enough, the number of markers $M$ has no effect on the variance). Across most of the parameter space, our expressions agree well with simulations. Notable deviations, however, arise for the SD in particularly short or long chromosomes. For these cases, the second, more complicated approximation, which we mentioned above and appears in Supplementary Section S1.3, is more accurate (Figure 2).

### 4. An approximate explicit expression

In this subsection, we derive another, simpler approximation of the variance, one that is less accurate but that has an explicit dependence on the population and genetic parameters. The gist of this approximation is that the main contribution to the variance comes from the long-distance probability of pairs of sites to reside on the same segment. Denote the distance between two given sites by $d$, and assume that $d > m$. For a given pair of individuals, if there was no recombination event between the two sites in the history of the two lineages, then both sites lie on a shared segment of length $\geq d > m$. Of course, even if there was a recombination event, the two sites could still be each on a different shared segment. However, this occurs with probability very close to $\pi^2$, the probability that the two sites are on shared segments given that they are independent.

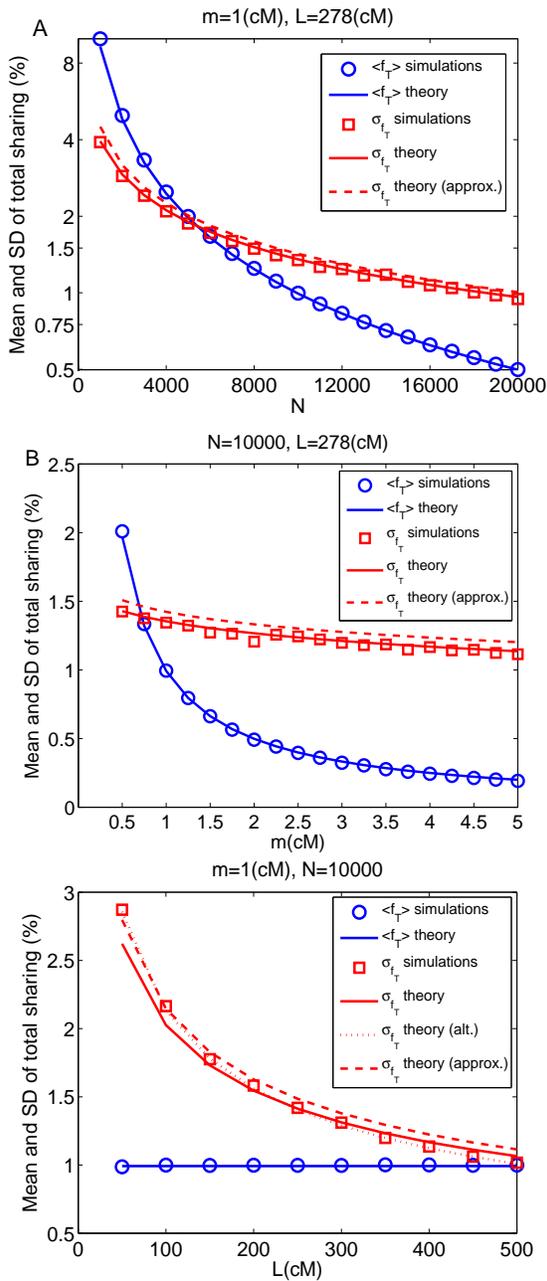

FIG. 2: **The mean and standard deviation of the total sharing.** For each parameter set, we used the GENOME coalescent simulator to generate a number of genealogies (from a population of size $N$ and for one chromosome of size $L$), and then calculated the lengths of IBD shared segments between random individuals. Each panel presents the results for the mean and standard deviation (SD) of the total sharing, that is, for each pair, the total fraction (in percentages) of the genome that is found in shared segments of length $\geq m$. Simulation results are represented by symbols, and theoretical results by lines (Eq. (4) for the mean and Eq. (12) for the SD are plotted in solid lines; the approximate form for the SD, Eq. (15), is shown in dashed lines). **A** We fixed $m = 1$cM and $L = 278$cM (the size of the human chromosome 1 [31]), and varied $N$. **B** Same as A, but with fixed $N = 10000$ and varying $m$. **C** Fixed $N$ and $m$ and varying chromosome length $L$. In this panel, we also plotted the result of an alternative, more elaborate calculation of the variance (dotted line; see Supplementary Section S1.3).

In terms of Eq. (6), the above approximation translates to, for $d > m$,

$$\pi_2(k) - \pi^2 \approx p_{\mathrm{nr}}, \tag{13}$$

where $p_{\mathrm{nr}}$ is the probability of no recombination,

$$p_{\mathrm{nr}} = \frac{1}{1 + \rho} = \frac{1}{1 + Nd/50} \approx \frac{50}{Nd} \tag{14}$$

for distant sites where $Nd \gg 50$. This is true because in the ancestral process (Figure 1), no recombination corresponds to a coalescence event taking place before any recombination event. Since the coalescence rate is 1 and the recombination rate is $\rho$, Eq. (14) follows. We can then further simplify and neglect the contribution to the variance from sites separated by short distance $d < m$. Finally, we can also neglect the single-site term of the variance, since it scales as $1/M$ and therefore vanishes when the markers are dense. Overall, the simplified Eq. (6) gives

$$\mathrm{Var}\left[f_T\right] \approx \frac{2}{M^2} \sum_{k=m\frac{M}{L}}^{M} (M - k) \frac{50}{Nk\frac{L}{M}} \tag{15}$$

$$\approx \frac{100}{MNL} \int_{m\frac{M}{L}}^{M} \frac{M - k}{k} dk = \frac{100}{NL} \int_{m/L}^{1} \frac{1 - x}{x} dx$$

$$= \frac{100}{NL} \left[\ln\left(\frac{L}{m}\right) - 1 + \frac{m}{L}\right] \approx \frac{100}{NL} \ln\left(\frac{L}{m}\right)$$

for $L \gg m$. Nicely, Eq. (15) provides an explicit (and rather simple) dependence on $N$, $L$ and $m$, and as expected, the expression does not depend on the marker density. Eq. (15) is also plotted in Figure 2, showing that it fits quite well to the simulation results, although it is usually less accurate than Eq. (12).

For the entire (autosomal) human genome, we use Eq. (5),

$$\mathrm{Var}\left[f_T\right] = \frac{100}{N} \frac{\sum_{i=1}^{22} L_i \ln\left(\frac{L_i}{m}\right)}{\left(\sum_{i=1}^{22} L_i\right)^2}.$$

For $m \approx 1$(cM), the last equation gives

$$\sigma_{f_T} \approx \frac{0.382}{\sqrt{N}}. \tag{16}$$

*A variable population size.* — The framework presented above can be extended to calculate the variance for a generalization of the Wright-Fisher model in which the population size is allowed to change in time. Denote the population size as $N(t) = N_0\lambda(t)$, where $t$ is the time (scaled by $N_0$) before present. The PDF of the (scaled) coalescence time for two lineages is (see, e.g., [32])

$$\Phi(t) = \frac{e^{-\int_0^t \frac{dt'}{\lambda(t')}}}{\lambda(t)}.$$

As shown in [10], the mean of the total sharing is obtained by substituting the above $\Phi(t)$ in Eq. (2), giving

$$\langle f_T \rangle = \int_0^\infty \Phi(t) \left(1 + \frac{mN_0 t}{50}\right) e^{-\frac{mN_0 t}{50}} dt. \quad (17)$$

For the variance, following Eq. (13), we need to calculate the probability of no recombination, $p_{\mathrm{nr}}$. For sites distance $d$ apart,

$$p_{\mathrm{nr}} = \int_0^\infty \Phi(t) e^{-\frac{tN_0 d}{50}} dt, \quad (18)$$

since for coalescence time $t$ the sites are separated by $2N_0 t$ meioses, in each of which the probability of no recombination is $e^{-d/100}$. Eq. (18) reduces to Eq. (14) for $\lambda(t) = 1$ (where $\Phi(t) = e^{-t}$). Eq. (18) can then be substituted into Eq. (15), giving

$$\mathrm{Var}\,[f_T] \approx \frac{2}{M^2} \sum_{k=m\frac{M}{L}}^{M} (M - k) \int_0^\infty \Phi(t) e^{-tN_0 k \frac{L}{M}/50} dt$$
$$\approx 2 \int_{m/L}^1 (1 - x) \left[\int_0^\infty \Phi(t) e^{-txN_0 L/50} dt\right] dx. \quad (19)$$

In Supplementary Section S1.5, we work out an example of a linearly expanding population, where Eq. (18) was solvable and the integral of Eq. (19) was evaluated numerically.

*The total sharing in a length range.*— Consider the quantity $f_{T;\ell_1,\ell_2}$, defined as the total fraction of the genome found in shared segments of length in the range $[\ell_1, \ell_2]$. Clearly, $f_{T;\ell_1,\ell_2} = f_{T;m=\ell_1} - f_{T;m=\ell_2}$, that is, the difference between the usual total sharing when $m = \ell_1$ and when $m = \ell_2$. The average is simply $\langle f_{T;\ell_1,\ell_2} \rangle = \langle f_{T;m=\ell_1} \rangle - \langle f_{T;m=\ell_2} \rangle = \frac{100N^2(\ell_2-\ell_1)[25(\ell_1+\ell_2)+\ell_1\ell_2 N]}{(50+\ell_1 N)^2(50+\ell_2 N)^2}$, an equation that was derived in [10] and then used for demographic inference. Here, we calculate the variance, $\mathrm{Var}\,[f_{T;\ell_1,\ell_2}]$, as follows,

$$\mathrm{Var}\,[f_{T;\ell_1,\ell_2}] = \mathrm{Var}\,[f_{T;m=\ell_1} - f_{T;m=\ell_2}] \quad (20)$$
$$= \mathrm{Var}\,[f_{T;m=\ell_1}] + \mathrm{Var}\,[f_{T;m=\ell_2}] - 2\mathrm{Cov}\,[f_{T;m=\ell_1}, f_{T;m=\ell_2}].$$

The covariance term can be expanded as

$$\mathrm{Cov}\left[\frac{1}{M}\sum_{s_1=1}^M I(s_1; m = \ell_1), \frac{1}{M}\sum_{s_2=1}^M I(s_2; m = \ell_2)\right]$$
$$= \frac{1}{M^2} \sum_{s_1} \sum_{s_2} \mathrm{Cov}\left[I(s_1; m = \ell_1), I(s_2; m = \ell_2)\right]$$
$$= \frac{1}{M^2} \sum_{s_1} \sum_{s_2} \left[\pi_2(s_1, \ell_1; s_2, \ell_2) - \pi_{m=\ell_1}\pi_{m=\ell_2}\right],$$

where $I(s; m = \ell)$ is the indicator that site $s$ is in a shared segment of length at least $\ell$, $\pi_{m=\ell}$ is the probability associated with the indicator, and $\pi_2(s_1, \ell_1; s_2, \ell_2)$ is the probability that site $s_1$ is in a shared segment of length at least $\ell_1$ and site $s_2$ is in a shared segment of length at least $\ell_2$. The key approximation, similar to the one made in subsection II A 4 (Eq. (15)), is that $\pi_2(s_1, \ell_1; s_2, \ell_2) - \pi_{m=\ell_1}\pi_{m=\ell_2}$ is non-zero only when the two sites lie on the same segment and the segment is *longer than* $\ell_2$. Defining $p_{\mathrm{nr}}$, as before, as the probability of no recombination between $s_1$ and $s_2$ in the history of the two individuals, we have

$$\mathrm{Cov}\,[f_{T;m=\ell_1}, f_{T;m=\ell_2}] \approx \frac{2}{M^2} \sum_{k=\ell_2 \frac{M}{L}}^{M} (M - k) p_{\mathrm{nr}}(k)$$
$$\approx \mathrm{Var}\,[f_{T;m=\ell_2}], \quad (21)$$

where the last step follows from Eq. (15). Substituting Eq. (21) into Eq. (20), we obtain

$$\mathrm{Var}\,[f_{T;\ell_1,\ell_2}] \approx \mathrm{Var}\,[f_{T;m=\ell_1}] - \mathrm{Var}\,[f_{T;m=\ell_2}].$$

For a constant population size, using Eq. (15) (taking all terms in that equation) gives

$$\mathrm{Var}\,[f_{T;\ell_1,\ell_2}] \approx \frac{100}{NL} \left[\ln\left(\frac{\ell_2}{\ell_1}\right) - \frac{\ell_2 - \ell_1}{L}\right]. \quad (22)$$

Eq. (22) is compared to simulations in Figure 3, showing good agreement. Note that as long as $\ell_1, \ell_2 \ll L$, the variance depends only on the ratio $\ell_2/\ell_1$.

### 5. The total sharing distribution and an error model

Having the first two moments of the total sharing, we sought to find its distribution, $P(f_T)$. While we could not find an exact expression, we could find, inspired by the numerical results of [13], a reasonable fit. Huff et al. [13] showed empirically that for HapMap's Europeans [31], the number of segments shared between random individuals was distributed as a Poisson, and that the length of each segment was distributed exponentially with a lower cutoff at $m$, independently of the number of segments. If this is true also for the Wright-Fisher model, then the total length of the shared segments, defined as $L_T = Lf_T$,

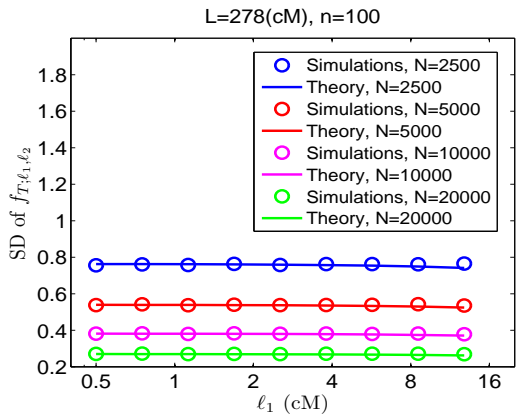

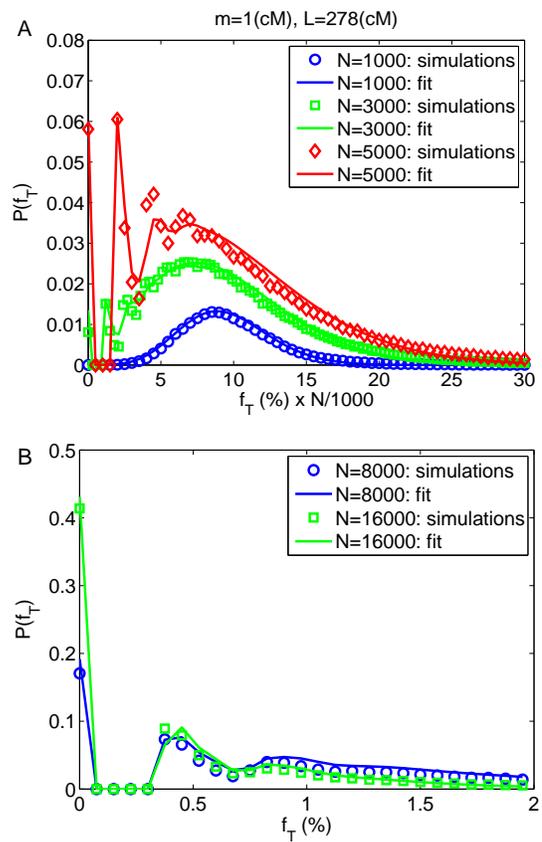

FIG. 3: **The standard deviation (SD) of the total sharing in a length range.** Simulation results (symbols) are shown for the SD of the fraction of the genome found in shared segments of specific length ranges. The total sharing for each range was calculated for random pairs of individuals in Wright-Fisher populations of the sizes indicated in the legend. The SD is plotted vs. the starting point of each length range, $\ell_1$ (where for each $\ell_1$, the successive data point is $\ell_2$). Note the logarithmic scale in the x-axis and hence that $\ell_2/\ell_1$ is fixed (equal to 1.5). Theory (lines) corresponds to Eq. (22).

is distributed as a sum of a Poisson distributed number of these exponentials. In equations,

$$P(L_T) = \sum_{n=0}^{\infty} e^{-n_0} \frac{n_0^n}{n!} \cdot \text{Prob}\{\ell_1 + \ell_2 + \ldots + \ell_n = L_T\}, \quad (23)$$

where $n_0$ is the mean number of segments, the density of the $\ell_i$s is $\exp[-(\ell_i - m)/\ell_0]/\ell_0$ ($\ell_0 + m$ is the mean segment length), and $\ell_i \geq m$. Such an expression is easier to handle in Laplace space, where the Laplace transform of $P(L_T)$, $\tilde{P}(s)$, is

$$\tilde{P}(s) = \sum_{n=0}^{\infty} e^{-n_0} \frac{n_0^n}{n!} \frac{e^{-mns}}{[s\ell_0 + 1]^n}$$

$$= \exp\left[-n_0\left(1 - \frac{e^{-ms}}{s\ell_0 + 1}\right)\right], \quad (24)$$

and we used the convolution theorem. For given $n_0$ and $\ell_0$, $P(L_T)$ (and then $P(f_T)$) was uniquely determined from $\tilde{P}(s)$ by numerical inversion [33, 34]. For specific values of $(N, L, m)$, we fitted the parameters $n_0$ and $\ell_0$ by minimizing the squared error between the simulated distribution and $P(f_T)$ (from Eq. (24)) in a grid search. The results are plotted in Figure 4, with the fitted $n_0$ and $\ell_0$ plotted in Figure **S2**. It can be seen that Eq. (24) captures quite well the unique features of $P(f_T)$ (except in the tail; see Figure **S2**).

Inspection of the distributions (Figure 4) for several values of $N$ lead to some interesting observations. For small $N$ (e.g., $N \approx 1000$, and for $m = 1$cM and

FIG. 4: **The distribution of the total sharing.** Simulation results (symbols) are shown for the distribution of the total sharing between random pairs of individuals in the Wright-Fisher model. Details of the simulation method are as in Figure 2A. **A** The distribution of the total sharing for $N = 1000$, 3000, and 5000. For better readability, the x-axis (the total sharing $f_T$) is given in percentages and scaled by $N/1000$, shifting the distributions for $N = 3000$ and $N = 5000$ to the right. **B** The distribution of the total sharing for $N = 8000$ and 16000. Here the x-axis is not scaled. In both panels, lines represent the fit to a sum of a Poisson number of shifted exponentials, Eq. (24).

$L = 278$cM), where the typical amount of sharing is large ($\langle f_T \rangle \approx 5 - 10\%$, $n_0 \approx 10$, $\ell_0 \approx 1$cM), the distribution is unimodal (but not normal), centered around $\langle f_T \rangle$. As $N$ increases (e.g., $N \approx 3000$), a discontinuous peak appears at $f_T = 0$, with $P(f_T) = 0$ for $0 < f_T < m/L (\approx 0.4\%)$. This is of course due to the restriction on the minimal segment length: a pair of individuals can share either nothing or at least one segment of length $m$. For $f_T > m/L$ the distribution is continuous, still centered around $\langle f_T \rangle$, but with small, yet notable peaks at $f_T = m/L, 2m/L, 3m/L, \ldots$ corresponding to pairs of individuals sharing a small number of minimal length segments. For even larger $N$ (e.g., $N \approx 10000$ and beyond), $\langle f_T \rangle$ drops below 1%, $n_0 \approx 1$ ($\ell_0$ still around 1cM), and the peaks at $f_T = 0$ and $f_T = m/L$ increase such that the distribution decreases almost monotonically beyond $m/L$. An analytical bound on the fraction of pairs not sharing any segment is given in Supplementary Section S2.1, Eq. S33.

*An error model.—* To model errors during IBD detection, suppose that we set $m$ large enough as to avoid any false positives (i.e., detected segments that are not truly IBD). We model false negatives as true IBD segments being missed with probability $\epsilon$ (independent of the segment length). It is possible to extend the above formulation (Eq. (23)) to the case with errors, as follows. Summing over the true number of segments, $n'$, the distribution of the number of detected segments, $n$, is

$$P(n) = \sum_{n'=n}^{\infty} e^{-n_0} \frac{n_0^{n'}}{n'!} \binom{n'}{n} (1-\epsilon)^n \epsilon^{n'-n}$$
$$= e^{-n_0(1-\epsilon)} \frac{[n_0(1-\epsilon)]^n}{n!},$$

that is, a Poisson with parameter $n_0(1-\epsilon)$. Then, as a sum of a random number of independent variables, the mean and variance of $L_T$ are $\langle L_T \rangle = \langle n \rangle \langle \ell \rangle$ and $\mathrm{Var}[L_T] = \langle n \rangle \mathrm{Var}[\ell] + \langle \ell \rangle^2 \mathrm{Var}[n]$, where $n$ is the number of segments and $\ell$ is the segment length. In our case,

$$\langle L_T \rangle = (1-\epsilon) n_0 (\ell_0 + m),$$
$$\mathrm{Var}[L_T] = (1-\epsilon) n_0 [\ell_0^2 + (\ell_0 + m)^2], \quad (25)$$

demonstrating that in the presence of detection errors, both the mean and the variance of the total sharing are $(1-\epsilon)$ times their noise-free values. This is confirmed by simulations in Figure 5.

*Other approaches.—* We note that a similar approach dates back to R. A. Fisher [35] and others [36–38] in their work on IBD sharing in a model where the population has been recently founded by a number of unrelated individuals. Briefly, those authors too assumed a Poisson number of IBD segments, each of which is exponentially distributed. They then matched the Poisson and exponential parameters to the average number of IBD sharing and the average number of segments, which they calculated using their population model. Here, we used a different population model (the coalescent; see Supplementary Section S2.2) and assumed the exponentials have a cutoff at $m$. In principle, the parameters $n_0$ and $\ell_0$ can also be directly calculated, by matching the mean and variance of the total sharing; see Supplementary Section S2.3. In practice, however, this does not give a good fit. In [10], a similar compound Poisson approach was developed but with a different, coalescent theory-based approximation of the segment length PDF, leading to an improved fit of the remaining parameter $n_0$.

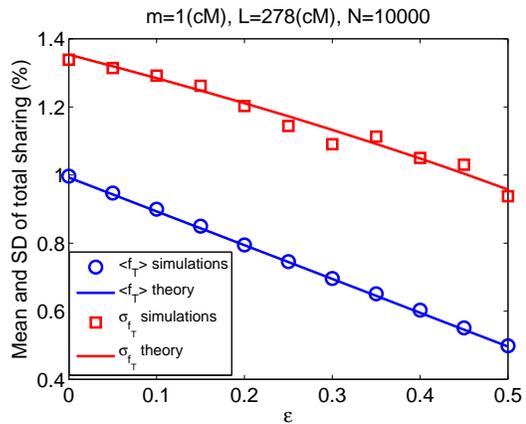

FIG. 5: **The mean and standard deviation (SD) of the total sharing in the presence of detection errors.** Simulation results (symbols) are plotted for mean and SD of the total sharing in the Wright-Fisher model. Simulation details are as in Figure 2, except that each segment was dropped with probability $\epsilon$. Theory (lines) is from Eq. (4) for the mean and (12) for the SD, but where the mean is multiplied by $(1-\epsilon)$ and the SD by $\sqrt{1-\epsilon}$, as in Eq. (25).

### B. The cohort-averaged sharing

We have so far considered the total sharing between any two random individuals in a population. In practice, we usually collect genetic information on a cohort of $n$ individuals. In this context, we can attribute each individual with the amount of genetic material it shares with the rest of the cohort. Define, for each individual, the *cohort-averaged* sharing $\overline{f_T}$, as the average total sharing between the given individual and the other $n-1$ individuals in the cohort. Naively, one may anticipate that the width of the distribution of $\overline{f_T}$ will approach zero for large $n$, because the averaging will eliminate any randomly arising differences between the individuals. We show that in fact, the width of the distribution approaches a non-zero limit. The individuals at the right tail of the cohort-averaged sharing distribution can be seen as "hyper-sharing", meaning they are, on average, more genetically similar to members of the cohort than are others. Similarly, individuals at the left tail are "hypo-sharing". The existence of hyper-sharing individuals is important for prioritizing individuals for sequencing, as we will show in Section II C.

Define the fraction of the genome shared by individuals $i$ and $j$ as $f_T^{(i,j)}$. The cohort-averaged sharing of $i$, $\overline{f_T}^{(i)}$, is

$$\overline{f_T}^{(i)} \equiv \frac{1}{n-1} \sum_{j=1, j \neq i}^{n} f_T^{(i,j)}.$$

The variance of $\overline{f_T}^{(i)}$ is

$$\mathrm{Var}\left[\overline{f_T}^{(i)}\right] = \frac{1}{(n-1)^2}\sum_{j=1, j\neq i}^{n}\mathrm{Var}\left[f_T^{(i,j)}\right]$$
$$+ \frac{1}{(n-1)^2}\sum_{j_1\neq i}\sum_{j_2\neq i, j_1}\mathrm{Cov}\left[f_T^{(i,j_1)}, f_T^{(i,j_2)}\right]$$
$$= \frac{\sigma_{f_T}^2}{n-1} + \frac{n-2}{n-1}\mathrm{Cov}\left[f_T^{(1,2)}, f_T^{(1,3)}\right]$$
$$\approx \frac{\sigma_{f_T}^2}{n} + \mathrm{Cov}\left[f_T^{(1,2)}, f_T^{(1,3)}\right], \qquad (26)$$

where we assumed $n \gg 1$ and used the fact that the covariance term is identical for all $(i, j_1, j_2)$ combinations and therefore, for simplicity of notation, we set $i = 1$, $j_1 = 2$, and $j_2 = 3$. Recall that $f_T^{(i,j)} = \frac{1}{M}\sum_{s=1}^{M}I(s)$ (Eq. (3)), where $I(s)$ is the indicator that site $s$ is on a shared segment. Thus, the covariance can be written as

$$\mathrm{Cov}\left[f_T^{(1,2)}, f_T^{(1,3)}\right] =$$
$$= \frac{1}{M^2}\sum_{s_1=1}^{M}\sum_{s_2=1}^{M}\left[\left\langle I^{(1,2)}(s_1)I^{(1,3)}(s_2)\right\rangle - \pi^2\right]$$
$$\approx \frac{2}{M^2}\sum_{k=1}^{M}(M-k)\left[\pi_2^{(1,2;1,3)}(k) - \pi^2\right],$$

where $I^{(i,j)}(s)$ is the indicator that site $s$ is on a segment shared between individuals $i$ and $j$, and $\pi_2^{(1,2;1,3)}(k)$ is the probability that a given site is on a segment shared between 1 and 2 and that another site, $k$ markers away from the first, is on a segment shared between 1 and 3. As in Section II A 4 (e.g., Eq. (15)), we will keep only the most dominant term in the sum. Consider the coalescent tree relating the three individuals 1, 2, and 3, and assume that the distance between the sites is $d > m$. If there was no recombination event in the entire tree between the two sites, then immediately $\pi_2^{(1,2;1,3)}(k) = 1$. Otherwise, we assume that each of the two sites belongs to a shared segment (or not) independently of the other, that is, $\pi_2^{(1,2;1,3)}(k) \approx \pi^2$. The probability of no recombination, $p_{\mathrm{nr}}$, depends on $T_3$, the total size of the tree of three lineages. Since the PDF of $T_3$ is $P(T_3) = e^{-T_3/2} - e^{-T_3}$ [24, 39],

$$p_{\mathrm{nr}} = \int_0^\infty P(T_3)e^{-dNT_3/100}dT_3 = \frac{5000}{(50+dN)(100+dN)},$$

or, for $dN \gg 100$,

$$p_{\mathrm{nr}} \approx \frac{5000}{(dN)^2}.$$

The covariance becomes

$$\mathrm{Cov}\left[f_T^{(1,2)}, f_T^{(1,3)}\right] \approx \frac{2}{M^2}\sum_{k=m\frac{M}{L}}^{M}(M-k)\frac{5000}{\left(k\frac{L}{M}N\right)^2}$$
$$\approx \frac{10000}{N^2L^2}\int_{m/L}^{1}\frac{1-x}{x^2}dx = \frac{10000}{N^2L^2}\left[\frac{L}{m}-1-\ln\left(\frac{L}{m}\right)\right]$$
$$\approx \frac{10000}{N^2mL}. \qquad (27)$$

Substituting Eqs. (15) and (27) in Eq. (26), the standard deviation of the cohort-averaged sharing is

$$\sigma_{\overline{f_T}} \approx \frac{\sigma_{f_T}}{\sqrt{n}}\sqrt{1 + \frac{n}{\sigma_{f_T}^2}\frac{10000}{N^2mL}}$$
$$\approx 10\sqrt{\frac{\ln\left(\frac{L}{m}\right)}{nNL}\left[1 + \frac{100n}{Nm\ln\left(\frac{L}{m}\right)}\right]}. \qquad (28)$$

For $(2\leq)\, n \ll Nm\ln\left(\frac{L}{m}\right)/100$, $\sigma_{\overline{f_T}} \approx \frac{\sigma_{f_T}}{\sqrt{n}}$, while for $n \gg Nm\ln\left(\frac{L}{m}\right)/100$ (but $< N$, as the size of the cohort size cannot exceed the population size), $\sigma_{\overline{f_T}} \approx \frac{100}{N\sqrt{mL}}$, which is independent of $n$. This implies that even for very large cohorts, the distribution of the cohort-averaged sharing retains a minimal width. Eq. (28) is in good agreement with simulations, as shown in Figure 6A (although some deviations are seen for larger $n$). We note that the variance was computed for a given individual over all ancestral processes of a cohort of size $n$. Therefore, the variance within the cohort, for a given ancestral process might actually be smaller. Simulations results (Figure S4), however, show that unless $n$ is very small, Eq. (28) is a good approximation also for the variance within the cohort.

For a genome with $c$ chromosomes,

$$\mathrm{Var}\left[\overline{f_T}\right] = \frac{\sum_{i=1}^{c}L_i^2\mathrm{Var}\left[\overline{f_{T,(i)}}\right]}{\left(\sum_{i=1}^{c}L_i\right)^2},$$

where $\overline{f_{T,(i)}}$ is the cohort-averaged sharing of chromosome $i$. For the human genome and for small $n$ and $m \approx 1\mathrm{cM}$, Eq. (16) gives

$$\sigma_{\overline{f_T}} \approx \frac{0.382}{\sqrt{nN}}, \qquad (29)$$

whereas for large $n$, Eq. (27) gives

$$\sigma_{\overline{f_T}} \approx \frac{1.68}{N\sqrt{m}}, \qquad (30)$$

which is, as explained above, independent of $n$.

The fact that the width of the cohort-averaged sharing distribution does not approach zero for large $n$ results from the "long-range" correlations between the averaged $(n-1)$ variables, or in other words, the fact that $\mathrm{Cov}\left[f_T^{(i,j_1)}, f_T^{(i,j_2)}\right] > 0$ for all $i, j_1, j_2$. In [40], it was found that when all pairs of random variables are weakly correlated, the PDF of their average converges to a normal distribution. In our case, the covariance is $\approx \frac{10000}{N^2 mL}$ (Eq. (27)), much smaller, for typical values of $N$, $L$, and $m$, than $\sigma_{f_T}^2 \approx \frac{100}{NL} \ln\left(\frac{L}{m}\right)$ (Eq. (15)). The variables we average are therefore weakly dependent, as we also observe in simulations (Figure **S5**). We thus conjectured that the distribution of the cohort-averaged sharing is normal or is close to one. This is confirmed by simulation results, as shown in Figure 6B. We note, however, that a small but systematic deviation from a normal distribution exists in all parameter configurations we tested, in the form of a broader right tail and a narrower left tail than expected (Figure **S5**). This seems to be due to the small probability ($\approx 1/N$) of random pairs of individuals to be close relatives.

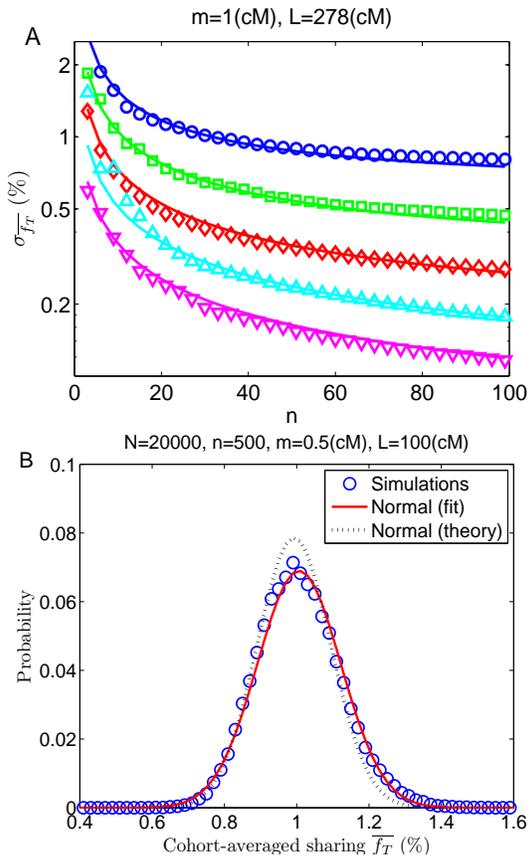

FIG. 6: **The cohort-averaged sharing.** **A** Simulation results (symbols) for $\sigma_{\overline{f_T}}$, that is, the standard deviation (SD) of the cohort-averaged sharing (in percentage of the chromosome) vs. the cohort size $n$. The different curves correspond to different values of $N$ (top to bottom: $N = 1000, 2000, 4000, 8000, 16000$). The lines correspond to Eq. (28). Details of the simulations are as in Figure 2A. **B** The distribution of the cohort-averaged sharing. The fit is to a normal distribution having the same mean and SD as the real data. Also plotted is a normal distribution with mean given by Eq. (4) and SD given by Eq. (28).

## C. Implications to sequencing study design

Suppose we have sparse genotype information for a large cohort, as well as whole-genome sequences for a subset of it. If the genotype data allow detection of IBD shared segments, then alleles not typed can be directly imputed if they lie on haplotypes shared with sequenced individuals (see, e.g., [22]). In fact, given abundant IBD sharing, such a strategy is expected to be quite successful; as we mentioned in Section II A 1, only about one recent mutation is expected on each shared segment. Since some individuals share more than others, their sequencing should be prioritized if imputation power is to be maximized. Recently, Gusev et al. [17] developed an algorithm (INFOSTIP) for sample selection based on the observed IBD sharing. Here, using our results in Section II B, we calculate the theoretical maximal imputation power.

Assume first that individuals are haploids; the case of diploids is treated later. Assume a cohort of size $n$, a budget that enables the sequencing of $n_s$ individuals, and two selection strategies: either of random $n_s$ individuals or of the $n_s$ individuals with the largest cohort-averaged sharing. Define an imputation metric, $p_c^{(i)}$, as the average fraction of the genome of $i$, an individual not sequenced, that is shared IBD with at least one sequenced individual. Let the selected individuals be $m_1, m_2, ..., m_{n_s}$, and denote the fraction of the genome that $i$ shares with $m_j$ as $f_T^{(i,m_j)}$. To calculate $p_c^{(i)}$, we assume that for all $j_1, j_2 = 1..., n_s$ ($j_1 \neq j_2$), $f_T^{(i,m_{j_1})}$ is independent of $f_T^{(i,m_{j_2})}$ (which is justified, as we showed in Section II B). We also assume that the locations of the shared segments are independent and uniformly distributed along the genome. Under these assumptions, the probability of a locus to be covered by at least one sequenced individual is

$$p_c^{(i)} = 1 - \prod_{j=1}^{n_s} \left(1 - f_T^{(i,m_j)}\right),$$ (31)

and this is also the average covered fraction of the genome. We note, however, that assuming that the locations of shared segments are independent and uniformly distributed is mostly for mathematical convenience. Simulation results (Figure **S6**) show that sharing tends to concentrate at specific loci, implying that Eq. (31) can be thought of as an upper bound (see Figure 7). When $f_T \ll 1$,

$$p_c^{(i)} \approx 1 - \exp\left[-\sum_{j=1}^{n_s} f_T^{(i,m_j)}\right],$$

and for random selection of individuals for sequencing,

$$p_c^{(\mathrm{rand})} \approx 1 - \exp\left(-n_s \left\langle f_T \right\rangle\right), \qquad (32)$$

where $\left\langle f_T \right\rangle$ is given by Eq. (4). When selecting the highest sharing individuals, values of $f_T^{(i,m_j)}$ come from the right tail of the cohort-averaged sharing distribution, $P(\overline{f_T})$,

$$\sum_{j=1}^{n_s} f_T^{(i,m_j)} \approx n_s \frac{\int_{\overline{f_c}}^{\infty} \overline{f_T} P(\overline{f_T}) d\overline{f_T}}{\int_{\overline{f_c}}^{\infty} P(\overline{f_T}) d\overline{f_T}} \equiv n_s \left\langle f_T^{(\mathrm{high})} \right\rangle,$$

where $\overline{f_c}$ and $\left\langle f_T^{(\mathrm{high})} \right\rangle$ are the minimum and average, respectively, of the cohort-averaged sharing among the sequenced individuals ($\int_{\overline{f_c}}^{\infty} P(\overline{f_T}) d\overline{f_T} = n_s/n$). Since we argued in Section II B that $P(\overline{f_T})$ is approximately normal with parameters $\left\langle f_T \right\rangle$ and $\sigma_{\overline{f_T}}$ (Eq. (28)), $\overline{f_c}$ satisfies

$$\mathrm{erfc}\left[\frac{\overline{f_c} - \left\langle f_T \right\rangle}{\sqrt{2}\sigma_{\overline{f_T}}}\right] = 2n_s/n. \qquad (33)$$

We can thus finally write

$$p_c^{(\mathrm{high})} \approx 1 - \exp\left(-n_s \left\langle f_T^{(\mathrm{high})} \right\rangle\right). \qquad (34)$$

Before getting to simulations, we notice that in practice, selection of *exactly* those individuals with the largest cohort-averaged sharing will not achieve the imputation power of Eq. (34). This is because the top sharing individuals could usually share many segments with each other and thus sequencing of all of them will be redundant (e.g., in the extreme case of siblings, both will appear as top sharing, but sequencing of both will add little power beyond sequencing just one). To avoid such redundancies, we selected the high-sharing (simulated) individuals using INFOSTIP [17], which proceeds in a greedy manner, each time selecting the individual who shares the most with the rest of the cohort in regions that are not yet covered by the already selected individuals. We then compared the imputation power when individuals were selected either randomly or using INFOSTIP. The results, shown in Figure 7, suggest good agreement between the theoretical Eqs. (32) and (34) and the simulations, at least as long as $n_s$ is not large (relative to $n$). For large $n_s$, the coverage is lower than predicted, likely due to the non-uniform concentration of the shared segments.

For a cohort of $n$ diploid individuals (assuming phase can be resolved) we redefine the cohort-averaged sharing as

$$\overline{f_{T,\mathrm{dip}}}^{(i)} \equiv \frac{1}{2}\left(\overline{f_T}^{(i,1)} + \overline{f_T}^{(i,2)}\right),$$

(where, e.g., $\overline{f_T}^{(i,1)}$ is the cohort-averaged sharing of the first chromosome of individual $i$) and assume that the individuals selected for sequencing have the largest diploid cohort-averaged sharing. Since the two terms in $\overline{f_{T,\mathrm{dip}}}^{(i)}$ are weakly dependent,

$$\sigma_{\overline{f_{T,\mathrm{dip}}}}(n) \approx \frac{1}{\sqrt{2}}\sigma_{\overline{f_T}}(2n),$$

where $\sigma_{\overline{f_T}}$ is given by Eq. (28). The coverage metric $p_c$ is interpreted, as before, as the probability of a locus on a given chromosome to be in a segment shared with at least one sequenced chromosome. The theory developed above is still valid, provided that in Eqs. (32) and (34) $n_s$ is replaced by $2n_s$ and that in Eq. (33) $\sigma_{\overline{f_T}}$ is replaced by $\sigma_{\overline{f_{T,\mathrm{dip}}}}$.

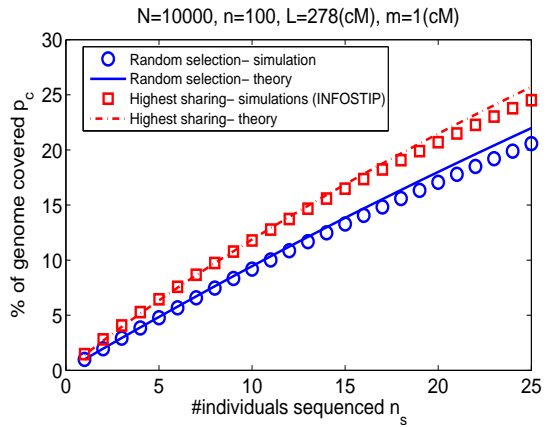

FIG. 7: **Coverage of genomes not selected for sequencing by IBD shared segments.** We simulated 500 Wright-Fisher populations with $N = 10000$, $n = 100$, and $L = 278$cM, and searched for IBD segments with length $\geq m = 1$cM. For each plotted data point, we selected $n_s$ individuals either randomly or using INFOSTIP. Then, for each of the $n - n_s$ individuals not selected, we calculated the fraction of their genomes shared with at least one selected individual. We plot (symbols) the average coverage over all individuals in all populations. Lines correspond to theory: Eq. (32) for random selection and Eq. (34) for INFOSTIP selection.

*Increase in association power.*— Using our results for the power of imputation by IBD, we calculate below the expected subsequent increase in power to detect rare variant association. We use the simple model of [41], in which we consider rare variants that appear in cases but not in any control, and assume that the causal variant is dominant.

Assume that we have genotyped and detected IBD segments in a cohort of $n_c$ (diploid) cases and $n_t$ controls, and that we sequenced a subset of $n_s$ individuals, out of which $n_{c,s}$ are cases and $n_{t,s}$ are controls ($n_s = n_{c,s} + n_{t,s}$). After imputation by IBD, a locus in a (diploid) individual not sequenced has probability $p_c^2$ to be successfully imputed, where $p_c$ is given by Eq (32) or Eq. (34). For a given locus, we define the *effective* number of cases (controls), as the number of cases (controls) for which genotypes are known either directly from sequencing or from imputation. Since there are $n_c - n_{c,s}$ cases not sequenced and $n_t - n_{t,s}$ controls not sequenced,

$$n_c^{(\text{eff})} \approx n_{c,s} + (n_c - n_{c,s})p_c^2(n_c, n_{c,s}),$$
$$n_t^{(\text{eff})} \approx n_{t,s} + (n_t - n_{t,s})p_c^2(n_t, n_{t,s}). \quad (35)$$

In the last equation we assumed, without loss of generality, that cases can only be imputed using other cases, and vice versa. The probability of a variant to appear in exactly $b$ cases but in no controls, under the null hypothesis that the variant assorts independently of the disease, is given by Fisher's exact test,

$$P(\text{cases only}) = \binom{n_c^{(\text{eff})}}{b} \Big/ \binom{n_c^{(\text{eff})} + n_t^{(\text{eff})}}{b}.$$

Define $Q$ as the threshold P-value and denote by $b^*$ the smallest integer above which $P(\text{cases only}) < Q$. When the causal variant carrier frequency in cases is $\beta$, the probability of the variant to appear in $b$ cases is binomial, and the power is, for a given $Q$,

$$\Pi = \sum_{b=b^*}^{n_c^{(\text{eff})}} \binom{n_c^{(\text{eff})}}{b} \beta^b (1-\beta)^{n_c^{(\text{eff})}-b}. \quad (36)$$

In Figure **S7**, we plot the power vs. $n_{c,s}$, when the sequencing budget $n_s = n_{c,s} + n_{t,s}$ is fixed and for representative parameter values. In Figure 8, we plot the power vs. the carrier frequency for the optimal value of $n_{c,s}$. The figure demonstrates that the power increases by several fold when imputation by IBD is used. This is, however, an expected consequence of increasing the effective sample size, and would likely be achieved with any imputation algorithm (e.g., [42]). The figure also shows an additional, slight increase when the highest sharing individuals are selected for sequencing. Thus, while it should be easy to identify the highest sharing individuals given a genotyped cohort (e.g., using INFOSTIP [17]), and doing so will increase the association power, our results suggest that the gain in power over a random selection will be minor.

### D. Other applications of the variance of IBD sharing

#### 1. An estimator of the population size

Assume that we have genotyped or sequenced a diploid chromosome of one individual and calculated $f_T$, the fraction of the chromosome shared between the individual's paternal and maternal chromosomes. Can we estimate the effective population size?

According to Eq. (4), $\langle f_T \rangle = \frac{100(25+Nm)}{(50+Nm)^2}$. Solving for $N$ gives (see also [10]),

$$N = \frac{50}{m \langle f_T \rangle} \left[ (1 - \langle f_T \rangle) + \sqrt{1 - \langle f_T \rangle} \right] \approx \frac{100}{m \langle f_T \rangle} - \frac{75}{m},$$

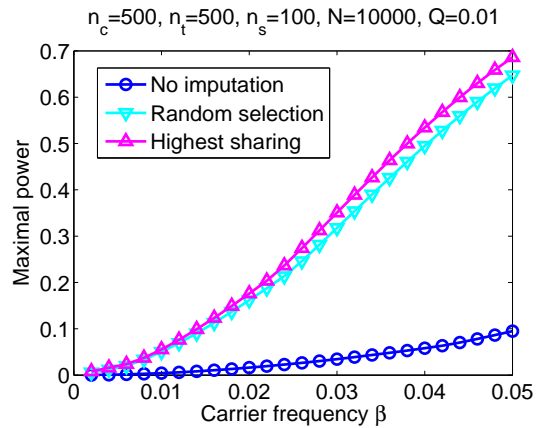

$n_c = 500,\ n_t = 500,\ n_s = 100,\ N = 10000,\ Q = 0.01$

FIG. 8: **Power to detect an association after imputation by IBD.** The figure shows the maximal power to detect an association, with and without imputation by IBD, and with sequenced individuals selected either randomly or according to their total sharing. The parameters we used were: $N = 10000$, $L = 278$cM (one chromosome), $m = 1$cM, cohort size of 500 cases and 500 controls, a total sequencing budget of $n_s = 100$ individuals, and a threshold P-value of $Q = 0.01$. For each carrier frequency $\beta$, we computed the power for each pair of $n_{c,s}$ and $n_{t,s}$ (number of sequenced cases and controls, respectively) such that $n_{c,s} + n_{t,s} = n_s$, and recorded and plotted the maximal power. The power was calculated using Eqs. (35) and (36), where in Eq. (35), $p_c$ was set to zero for the case of no imputation, or calculated using Eqs. (32) and (34) (random selection and selection by total sharing, respectively, and adjusted for diploid individuals). For the studied parameter set, imputation by IBD leads to a major increase in power. Proper selection of individuals for sequencing also contributes to the power but only slightly.

for $\langle f_T \rangle \ll 1$. This suggests the following estimator,

$$\hat{N} = \frac{100}{m f_T} - \frac{75}{m}. \quad (37)$$

Below, we investigate the properties of the simple estimator of Eq. (37). Using Jensen's inequality, it is easy to see that the estimator is biased,

$$\langle \hat{N} \rangle = \frac{100}{m} \left\langle \frac{1}{f_T} \right\rangle - \frac{75}{m} \geq \frac{100}{m \langle f_T \rangle} - \frac{75}{m} = N.$$

The variance of $\hat{N}$ is proportional to $\text{Var}\,[1/f_T]$, which we could not calculate, but could approximate as follows. Let us write $\hat{N}$ as

$$\hat{N} = \frac{100}{m[(f_T - \langle f_T \rangle) + \langle f_T \rangle]} - \frac{75}{m}$$
$$\approx \frac{100}{m \langle f_T \rangle} \left( 1 - \frac{f_T - \langle f_T \rangle}{\langle f_T \rangle} \right) - \frac{75}{m},$$

where we applied the Taylor expansion $\frac{1}{1+x} \approx 1 - x$, assuming $|f_T - \langle f_T \rangle| \ll \langle f_T \rangle$ (in which regime clearly

$\langle \hat{N} \rangle = N$). Since additive constants do not contribute to the variance, the standard deviation is

$$\sigma_{\hat{N}} \approx \frac{100 \sigma_{f_T}}{m \langle f_T \rangle^2} \approx \frac{mN^{3/2}}{10} \sqrt{\ln\left(\frac{L}{m}\right)},$$

where we used $\langle f_T \rangle \approx 100/(mN)$ (Eq. (4)), and Eq. (15)) for $\sigma_{f_T}$. The effective population size can also be inferred using Watterson's estimator, which is $\hat{N}_W = S_2/(2\mu)$, where $S_2$ is the number of heterozygous sites and $\mu$ is the mutation rate (per chromosome per generation). Watterson's estimator is unbiased, $\langle \hat{N}_W \rangle = N$, and has variance (assuming no recombination) $\mathrm{Var}\left[\hat{N}_W\right] = [2\mu N + (2\mu N)^2]/4\mu^2 \approx N^2$. Therefore, $\sigma_{\hat{N}_W}/N \approx 1$, compared to $\sigma_{\hat{N}}/N \approx N^{1/2}$ for the IBD estimator.

Note that in practice, the proposed estimator is not very useful, as it diverges whenever $f_T = 0$ (which is common for large $N$). Suppose, however, that we have sequences for $n$ (haploid) chromosomes, and that we have computed the total sharing between all pairs. Define $\overline{\overline{f_T}} = \sum_i \sum_{j>i} f_T^{(i,j)}/\binom{n}{2}$. The estimator now takes the form

$$\hat{N} = \frac{100}{m\overline{\overline{f_T}}} - \frac{75}{m}. \tag{38}$$

This is again an overestimate, $\langle \hat{N} \rangle \geq N$. In Supplementary Section S3, we show that $\sigma_{\hat{N}}$ is approximately

$$\sigma_{\hat{N}} \approx \frac{mN^{3/2}}{5\sqrt{nL}} \sqrt{\frac{\ln\left(\frac{L}{m}\right)}{2n} + \frac{100}{Nm}}, \tag{39}$$

For comparison, in Watterson's estimator for $n$ (haploid) chromosomes, $\sigma_{\hat{N}_W}/N \approx 1/\ln n$ (for large $N$ and $n$), which decays to zero with increasing $n$ slower than the IBD estimator. Simulation results, shown in Figures **S8** and **S9**, confirm the accuracy of Eq. (39) and show that the bias is limited to very small values of $n$.

In the context of the error model of Section II A 5, introducing a probability $\epsilon$ to miss a true IBD segment will decrease the average total sharing by $(1-\epsilon)$ (Eq. (25)). Consequently, Eq. (38) will estimate a population size approximately $1/(1-\epsilon)$ ($\approx (1+\epsilon)$ for small $\epsilon$) larger than the true one.

### 2. IBD sharing between siblings

The total IBD sharing between relatives can usually be decomposed into sharing due to the recent co-ancestry and "background" sharing due to population inbreeding [13, 14]. While much is known about the distribution of sharing in pedigrees (e.g., [43]), less is known about the population level sharing, and relatedness detection algorithms (e.g., [13, 14]) estimate it empirically. In a different domain, the variance in sharing between relatives appears in theoretical calculations of the variance of heritability estimators [44]. Our results for the variance of the total sharing in the Wright-Fisher model (Section II A) can thus have practical applications if modified to account for recent co-ancestry.

Here, we calculate the variance of the sharing between siblings by combining the approach of [44] with that of our Section II A 4. Assume that two individuals are siblings, either half or full— we will calculate, without loss of generality, only the sharing between the two chromosomes that descended from the same parent— and denote the fraction of sharing as $f_S$. Assume as before a population of size $N$ and one chromosome of length $L$. For a given marker to be on a shared segment, it can either be on a segment directly co-inherited from the same grandparent (probability $1/2$), or otherwise on a segment shared between the grandparents (probability $\pi/2$, Eq. (2)). We ignore boundary effects near the sites of recombination at the parent. The mean fraction of the genome shared is therefore just $\langle f_S \rangle = (1 + \pi)/2$. The variance can be written as in Eq. (6),

$$\mathrm{Var}\left[f_S\right] \approx \frac{2}{M^2} \sum_{k=1}^{M} (M-k) \left[\pi_{2,S}(k) - \frac{1}{4}\left(1+\pi\right)^2\right],$$

where $\pi_{2,S}(k)$ is the probability of two sites separated by $k$ markers (or genetic distance $d = k\frac{L}{M}$) to be on segments shared between the siblings. The probability that the two sites are both co-inherited from the same grandparent is

$$p_{\mathrm{same}} = \frac{1}{2}\left[(1-r)^2 + r^2\right] = \frac{1}{4}\left(1 + e^{-d/25}\right),$$

where $r$ is the recombination fraction and we used Haldane's map function [44]. Also with probability $p_{\mathrm{same}}$, the sites are both inherited from different grandparents, and we use the expressions developed in Section II A 4 for the probability of the sites to be in shared segments: $\pi_{2,S}(k) = \pi^2 + \Theta(d-m)p_{\mathrm{nr}}$ (where $p_{\mathrm{nr}} \approx 50/(Nd)$ and $\Theta(x) = 1$ for $x > 0$ and is zero otherwise). With probability $(1 - 2p_{\mathrm{same}})$, one site is co-inherited and the other is not; in that case $\pi_{2,S}(k) = \pi$. Approximating the sum as an integral and simplifying, we finally have

$$\mathrm{Var}\left[f_S\right] \approx 2 \int_0^1 dx(1-x) \left\{ \frac{\pi\left(1 - e^{-xL/25}\right)}{2} + \right. \tag{40}$$

$$\left. \frac{1 + e^{-xL/25}}{4}\left[1 + \pi^2 + \Theta\left(x - \frac{m}{L}\right)\frac{50}{NxL}\right] - \frac{(1+\pi)^2}{4} \right\}.$$

We solved Eq. (40) using MATHEMATICA and summed over all chromosomes as in Eq. (5). The results for the mean and standard deviation (SD) of the total sharing between siblings are plotted in Figure 9 and compared to an outbred population where the grandparents are unrelated. The SD in the outbred population overestimates the Wright-Fisher SD, in up to $\approx 18\%$ for $N$ as small as 500.

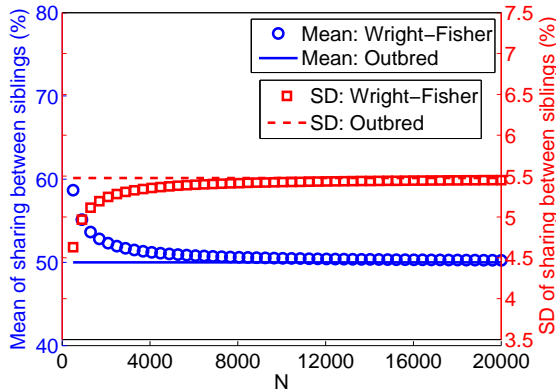

FIG. 9: **IBD sharing between siblings in the Wright-Fisher model**. We plot the theoretical mean and standard deviation (SD) of the IBD sharing between the (maternal only or paternal only haploid) genomes of siblings. Lines correspond to an outbred population (unrelated grandparents): the mean sharing is 50% and the SD is taken from [44]. Symbols correspond to the theory for Wright-Fisher model: the mean sharing is $(1 + \pi)/2$ (where $\pi$ is given by Eq. (2)), and the SD is given by Eq. (40). We used $m = 1\text{cM}$ and the chromosome lengths of the autosomal human genome. Note that the y-axis is on the left side for the mean and on the right side for the SD.

### 3. IBD sharing after an admixture pulse

In this final subsection, we study the IBD sharing in a simple admixture model. In our model, a single population $A$ of constant size $N$ has received gene flow from population $B$, $G_a$ generations ago. We assume that gene flow took place for one generation only (hence, an admixture pulse), and further, that population $B$ is sufficiently large that the chromosomes it donated to $A$ share no detectable IBD segments. Denote the fraction of the lineages coming from population $A$ at the admixture event as $\alpha$ (a fraction $1 - \alpha$ coming from $B$), and let $T_a = G_a/N$ be the scaled admixture time. We are interested in IBD sharing between extant chromosomes in population $A$.

To approximate the mean IBD sharing in the sample, note that if admixture was very recent, then two chromosomes will be potentially shared only if both descend from population $A$, which occurs with probability $\alpha^2$. Therefore, the mean sharing is $\alpha^2$ times its value without admixture. While this is a good approximation (Figure **S10**), it does not account for two chromosomes, one or two of which are from the external population $B$, having their common ancestor more recently than the admixture event. We therefore calculate the mean IBD sharing using Eq. (17), using the following (non-normalized) PDF

for the coalescence times:

$$\Phi(t) = \begin{cases} e^{-t} & t < T_a, \\ \alpha^2 e^{-t} & t > T_a, \end{cases} \tag{41}$$

which gives

$$\langle f_T \rangle = \int_0^\infty \Phi(t) \left(1 + \frac{mNt}{50}\right) e^{-\frac{mNt}{50}} dt$$
$$= \alpha^2 \frac{100(25 + mN)}{(50 + mN)^2} + T_a(1 - \alpha^2) + \mathcal{O}(T_a^2). \tag{42}$$

Note that this is just $\langle f_T \rangle_{\text{admix}} \approx \alpha^2 \langle f_T \rangle_{\text{no admix}} + (1 - \alpha^2)T_a$. The first term corresponds to lineages descending from population $A$; the second term corresponds to at least one of the lineages descending from population $B$ but where the lineages have coalesced already in the hybrid population. The variance can be similarly calculated, by substituting Eq. (41) into Eq. (19),

$$\text{Var}\left[f_T\right] \approx 2 \int_{m/L}^1 (1-x) \left[\int_0^{T_a} e^{-t - txNL/50} dt\right] dx$$
$$+ 2\alpha^2 \int_{m/L}^1 (1-x) \left[\int_{T_a}^\infty e^{-t - txNL/50} dt\right] dx$$
$$\approx \frac{100}{NL} \left\{\ln\left(\frac{L}{m}\right) - 1 + (1 - \alpha^2)\left[\gamma - \left|\ln\left(\frac{mNT_a}{50}\right)\right|\right]\right\}, \tag{43}$$

where $\gamma$ is the Euler-Mascheroni constant, and we solved the integrals in MATHEMATICA and later simplified under specific assumptions (see Supplementary Section S4). Eq. (43) usually predicts a variance slightly smaller than the case of no admixture. Simulation results are shown in Figure **S10** for the mean and variance. While agreement is not perfect (as Eq. (19) is itself approximate), Eqs. (42) and (43) capture the main effects of changing $\alpha$ and $T_a$. Note that the result of Eq. (42) implies that, for small $T_a$ and large $N$, the observed mean IBD sharing is as if the population is of size $\approx N/\alpha^2$.

*A test for admixture.* —. For recent admixture (small $T_a$), the fractions of ancestry vary among individuals [45, 46]. In our model, since a pair of segments is shared mostly when both descend from population $A$, some individuals will share more than others merely due to having a larger fraction of $A$ ancestry. In turn, this will increase the variance of the cohort-averaged sharing. This observation suggests the following test for a recent gene flow into a population. *(i)* Extract IBD segments and calculate the mean fraction of total sharing over all pairs, $\overline{\overline{f_T}}$, as well as the SD of the cohort-averaged sharing, $\sigma_{\overline{f_T}}$. *(ii)* Use Eq. (38) to infer the population size: $\hat{N} = 100/(m\overline{\overline{f_T}}) - 75/m$. *(iii)* Simulate $N_{\text{pop}}$ populations of size $\hat{N}$; extract IBD sharing and calculate the SD of the cohort-averaged sharing in each population. *(iv)* The P-value for rejecting the null hypothesis of no admixture is the fraction of the $N_{\text{pop}}$ populations where the SD of the cohort-averaged sharing was larger than the observed one. Note that the identity of the external population need not be known, nor are the admixture fraction and time; the test relies on admixture creating a gradient of ancestry fractions and hence an increased variability in the similarity between individuals. Simulation results are plotted in Figure **S10**, showing that for a P-value of 0.05 and $G_a = 5$, gene flow with $\alpha \approx 0.9$ or lower can be detected ($\alpha \approx 0.8$ or lower for $G_a = 10$). We stress that a broader than expected distribution of cohort-averaged sharing does not *necessarily* indicate admixture, and there might be other factors responsible for the effect (see also the Discussion). We validated, however, that IBD detection errors alone (as in the model of Section II A 5) as well as variable population size (in a simple two-size model) do not lead to significant P-values in the admixture test (Figure **S11**).

*IBD sharing and admixture in the Ashkenazi Jewish population.*— As our final result, we apply the admixture test to the real population of Ashkenazi Jews (AJ). Historical records, and recently also genetic studies, suggest that AJ form a genetically distinct group of likely Middle-Eastern origin. However, the AJ population was also shown to receive a significant amount of gene flow from neighboring European populations [47–51]. We analyzed a dataset of $\approx 2600$ AJ, details on which have been published elsewhere [10, 51] and are summarized in the Methods section. To detect IBD shared segments in the AJ population, we used GERMLINE [2]. For 500 (diploid) individuals on chromosome 1, and with $m = 1\mathrm{cM}$, the average fraction of sharing over all pairs is $\approx 4\%$, leading to an estimated population size of $\hat{N} \approx 4500$ (diploids). The SD of the cohort-averaged sharing is 0.52%, higher than the SD in all 500 populations we simulated with a constant size $\hat{N}$ (typically 0.27%, maximum 0.30%). The recent history of Ashkenazi Jews, however, has likely involved bottlenecks and expansions, different from the constant size assumption. In [10], a population model was inferred based on the fraction of the genome shared at different segment lengths. The model's best estimate of AJ history is a slow expansion until about 35 generations ago, and then a severe bottleneck (effective population size of just 270) followed a by rapid expansion to a current size of a few millions. As can be seen in Figure 10A and B, the model agrees well with the distribution of the fraction of total sharing over all pairs, but predicts a much narrower distribution of cohort-averaged sharing than the true one. Here too, in none of 100 simulated populations with the inferred demography was the SD of the cohort-averaged sharing as large as in the real data. These results, therefore, suggest (based on the AJ population alone) that the AJ population was the target of a recent gene flow. To confirm that the increase in the variance of the cohort-averaged sharing is due (at least partly) to admixture, we ran an admixture analysis (ADMIXTURE [52]) comparing AJ to HapMap's CEU [31]. As can be seen in Figure 10C, the fraction of "AJ ancestry" is indeed highly correlated with the cohort-averaged sharing (Pearson's $r = 0.59$).

## III. DISCUSSION

The recent availability of dense genotypes, together with sophisticated detection tools, have transformed IBD sharing into an increasingly important tool in population genetics. Here, we used coalescent theory to compute the variance and other properties of the total sharing in the Wright-Fisher model. For the variance, we suggested three derivations, one of which was more coarse but had a simple closed form that was later extended to populations of variable size. Investigating the cohort-averaged sharing, we discovered the curious phenomenon of 'hyper-sharing'. We showed how this can be exploited to improve power in imputation and association studies. We also calculated the variance of the total sharing between siblings, and briefly considered some implications to the accuracy of demographic inference. We finally investigated IBD sharing in a hybrid population and suggested a test for admixture based on the cohort-averaged sharing, which we then applied to the Ashkenazi Jewish population. We provide MATLAB routines for the main results.

Most of our analytical results depend on certain assumptions and simplifications, as specified in the individual sections and in Supplementary Section S1.2. Additionally, in reality, the Wright-Fisher model and the coalescent are only approximations of the true ancestral process, and procedures such as phasing, IBD inference, and imputation are also prone to error. IBD detection errors will particularly affect our results for imputation and association studies (Section II C), and these results should therefore be considered as idealized upper bounds. The error model we introduced, where each IBD segment is missed with a certain probability, gives a sense of the effect of errors. Investigation of more detailed models, e.g., length dependent error rate for segment misdetection or more realistic models for imputation and association studies, is challenging and left for future work.

Prospects of our work are in a few fields. First, as shown in [10], theoretical characterization of IBD sharing can lead to new methods for demographic inference, which are expected to perform particularly well when investigating the recent history of genetic isolates. Here, we expanded the theory of IBD sharing to compute the variance of the total sharing and the cohort-average sharing. This turned useful, for example, when we provided in Section II D 1 expressions for the variance of an estimator of the population size based on the average sharing over all pairs of chromosomes and in Section II D 3, where we described a test for recent admixture. In another domain, understanding the distribution of sharing between relatives can improve the accuracy of relatedness detection (Section II D 2). Other potential applications are in the detection of regions either positively selected or associated with a disease based on excess sharing, although more work is needed for these. Finally, our results provide the first estimate for the potential success of imputation by IBD strategies (Section II C). We note that of course, once a given cohort has been genotyped, IBD can be calculated directly to estimate the expected success of imputation. However, in many cases, study design takes place before the actual recruiting and genotyping, and then, if a rough estimate of the population size is available, our results can be invoked to estimate the amount of resources needed.

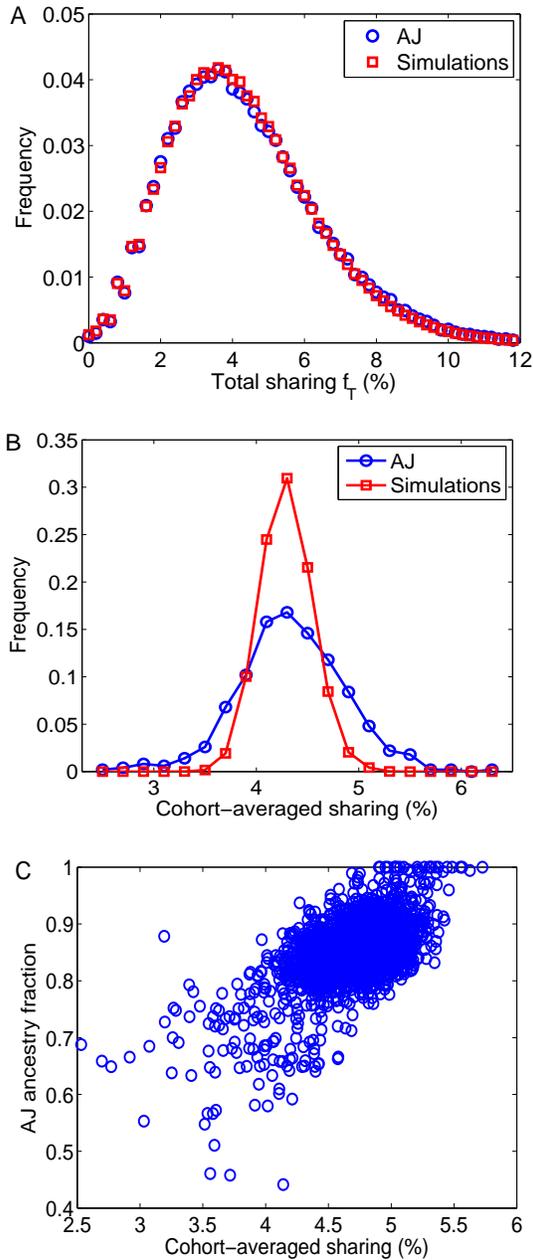

FIG. 10: **IBD sharing and admixture in the Ashkenazi Jewish (AJ) population.** We detected IBD shared segments using GERMLINE in chromosome 1 of $n = 500$ AJ individuals and compared to simulations of the demographic history inferred in [10]. **A** The distribution of the total sharing over all pairs. **B** The distribution of the cohort-averaged sharing. While the demographic model fits well the sharing distribution over all pairs, the distribution of the real cohort-averaged sharing is broader than in the model. **C** We used ADMIXTURE to calculate the admixture fraction of AJ individuals compared to the CEU population. The "AJ ancestry fraction" of each individual is plotted against its cohort-averaged sharing. This panel shows results for the full dataset ($\approx 2600$ individuals).

One of our interesting findings was the presence of hyper-sharing individuals. While we did not define the term precisely, we referred to the fact that even for large cohorts, the variance of the cohort-averaged sharing does not decrease below a certain value. This result, while somewhat counterintuitive, follows naturally from the population model. In the real population of Ashkenazi Jews (AJ), we showed that the distribution of the cohort-averaged sharing is even broader, indicating possible admixture, and indeed, we found that the cohort-averaged sharing is highly correlated with the Ashkenazi ancestry fraction. This is not to say that admixture was the only factor shaping the distribution of IBD sharing; other factors such as selection or population substructure could have been playing a role as well. Our results, however, emphasize the importance of reconstructing the AJ demography simultaneously with that of their neighboring populations.

## IV. METHODS

*Coalescent simulations.* — To simulate IBD sharing in the Wright-Fisher model, we used the GENOME haploid coalescent simulator [30]. Recombination in GENOME is discretized to short blocks and mutations (which we ignore in this study) are placed on the simulated branches. In all simulations, we generated one chromosome with recombination rate of $10^{-8}$ per generation per base-pair and block lengths of $10^4$ base-pairs (corresponding to resolution of 0.01cM in the lengths of the shared segments).

*IBD sharing in simulations.* — We used an add-on to GENOME that returns, for each pair of chromosomes, the locations of all shared segments [10]. In that add-on, a segment is shared as long as the two chromosomes share the same ancestor, even if there was a recombination event within the segment. We calculated, for each pair, the total length of shared segments longer than $m$, and divided by the chromosome size. For Figures 2-6, we simulated $N_{pop} \geq 100$ populations and $n = 100$ haploid sequences in each population, and calculated all properties of the total sharing among all $N_{pop}\binom{n}{2}$ available pairs. For the cohort-averaged sharing, we averaged, for each of the $n$ chromosomes, their sharing to each of the other $n-1$ chromosomes in the cohort, and then used the $N_{pop}n$ calculated values to obtain the variance and the distribution. Details on the simulation of an admixture pulse can be found in Supplementary Section S4.

*The Ashkenazi Jewish (AJ) cohort.* — The cohort we analyzed was previously described in [10, 51]. Briefly, DNA samples from $\approx$ 2600 AJ were genotyped on Illumina-1M SNP array. Genotypes (autosomal only) were subjected to quality control, including removal of close relatives, and phasing (Beagle [53]), leaving finally $\approx$ 741,000 SNPs for downstream analysis. IBD sharing was calculated using Germline [2] with the following parameters: bits: 25, err_hom: 0, err_het: 2, min_m: 1, h_extend: 1. The results presented in Section II D 3 remained qualitatively the same even when we used a longer length cutoff of $m = 5$cM.

*Admixture analysis.* — For the admixture analysis, we merged the HapMap3 CEU population (Utah residents with ancestry from Northern and Western Europe) [31] (release 2) with the AJ data, removed all SNPs with potential strand inconsistency, and pruned SNPs that were in linkage disequilibrium [5]. We then ran Admixture [52] with default parameters and $K = 2$. Admixture consistently classified all individuals according to their population (CEU/AJ). Genome-wide, the AJ ancestry fraction was $\approx$ 85%, compared to $\approx$ 3% for the CEU population. Principal components analysis (SmartPCA

[54]) gave qualitatively similar results.

*Simulations of AJ demography.* — Demographic reconstruction of the AJ population was performed in [10] using chromosome 1 of 500 randomly selected individuals and using a novel IBD-based method described therein. Simulations presented here were performed using the final set of inferred demographic parameters: ancestral (diploid) population effective size of $\approx$ 2300 individuals, expansion starting 200 generations ago reaching $\approx$ 45000 individuals 33 generations ago, a severe bottleneck of $\approx$ 270 individuals, and an expansion to the current size of $\approx$ 4.3 million individuals. Simulation of one hundred populations was carried out using Genome [30].

## V. ACKNOWLEDGEMENT

We thank the reviewers for insightful comments and Omer Bobrowski for discussions. S. C. thanks the Human Frontier Science Program for financial support. I. P. acknowledges support from NSF grant CCF 0845677 and NIH grant U54 CA121852 .

# Extended analytical results

## S1 The variance of the total sharing

### S1.1 The expected number of recent mutations on a shared segment

Consider a segment shared IBD between two individuals. Regardless of the segment length, the two individuals are expected to differ in $\approx 1$ site along the segment. This is because for a pair of individuals with MRCA $g$ generations ago, the shared segment is of typical length $100/(2g)$ cM (see, e.g., *Mean total sharing* section in the main text). The number of recent mutations per cM is $2g\mu$, where $\mu$ is the mutation rate per generation per cM. The total number of differences is therefore approximately

$$\# \text{ differences} \approx \frac{100}{2g}2g\mu = 100\mu. \tag{1}$$

For the human genome, $\mu \approx 10^{-8}$ per generation per bp [1], or $\approx 0.8 \cdot 10^{-2}$ per generation per cM (1MB corresponds roughly to 1.25cM). The number of difference is therefore around 1.

### S1.2 The assumptions underlying derivation of the variance of the total sharing

We summarize below the assumptions made when calculating the mean and the variance of the total sharing (main text *Mean total sharing* and *The variance of the total sharing* sections).

1. The population is Wright-Fisher with constant (effective) size $N$. We do not distinguish between male and female history, and all present-day individuals are represented as random pairs of haploids from the current generation.

2. The ancestral process is described by Kingsman's coalescent [2]; specifically, time is assumed to be continuous, and the distribution of coalescence times is exponential with rate 1.

3. Recombination is a Poisson process with rate 0.01 per cM.

4. The recombination rate between markers is proportional to the genetic distance between the them.

5. The markers are equally spaced, in genetic distance, along each chromosome and are dense enough, that when calculating the probability that a segment has length $\geq m$, we can ignore the discreteness of the markers.

6. If two sites are on different chromosomes, they are shared or not independently of each other.

7. Boundary effects at the ends of the chromosomes are ignored.

8. We assume that the events that two sites are in shared segments are independent once we specify the time to the MRCA at each site.

Assumptions 1, 2, 3, and 4 are standard when studying finite, isolated populations [2]. Assumption 5 should present no problem in practice, with SNP arrays covering over a million sites or with whole-genome sequences. For assumption 6, we can, approximately, expect segments on different chromosomes to be shared independently of each other if the individuals are sufficiently unrelated that the average number of segments shared genome-wide is less than one, which is true for 4th (half-) cousins or less related individuals [3]. Assumption 7 is reasonable when $L \gg m$ ($L$ is the length of the chromosome, $m$ is the minimal segment length).

 S. Carmi *et al.*

For the last assumption (8), one may suggest that if there was no recombination event in the history of two sites, then they are not independent. The reason why our approximation works is that when the two sites have the same coalescence time, it is usually very short (otherwise there was been a recombination event and the coalescence times would not be the same in the two sites), increasing the probability that they lie on shared segments. If the sites have different coalescence times, the times would tend to be longer, reducing the probability that the sites are on shared segments, in accordance with the fact that they were separated by a recombination event.

One importance of the derivation presented in the main text is that it sets the framework for a more detailed calculation that eliminates the last assumption. It does so by conditioning the probability $\pi_2(s_1, s_2)$ on whether or not there was a recombination event. For each case, it then proceeds using the Markov chain representation of coalescent with recombination. This is explained in the next subsection.

## S1.3  An alternative calculation of the variance of the total sharing

In this subsection, we recalculate the probability $\pi_2(s_1, s_2) = \pi_2(k)$ of two sites separated by $k$ markers to be both on shared segments of length $\geq m$. We use the Markov chain illustrated in Figure 1 of the main text as well as other notation as used in the main text. As mentioned above, we calculate $\pi_2$ by conditioning on whether or not the two sites have been separated by a recombination event,

$$\pi_2 = p_{\text{nr}}\pi_{\text{nr}} + (1 - p_{\text{nr}})\pi_{\text{r}}, \tag{2}$$

where $p_{\text{nr}}$ is the probability of no recombination, $\pi_{\text{nr}}$ is the probability of both sites to be in shared segments when there was no recombination, and $\pi_{\text{r}}$ is the probability of both sites to be in shared segments when there was recombination.

To calculate the probability of no recombination, we consider the discrete time Wright-Fisher model (as we found that it matches better the discrete-time simulations). In discrete time, the PDF of $g$, the number of generations to the (single-site) MRCA, is geometric, $P(g) = \frac{1}{N}\left(1 - \frac{1}{N}\right)^{g-1}$. Given an MRCA at generation $g$ at one site, we require that there was no recombination between that site and the other site, in both chromosomes, and in all $g$ generations. Because recombination is a Poisson process and the distance between the sites is $d = k\frac{L}{M}$, there will be no recombination with probability

$$p_{\text{nr}} = \sum_{g=1}^{\infty} \frac{1}{N}\left(1 - \frac{1}{N}\right)^{g-1} e^{-dg/50} = \frac{1}{1 + N\left(e^{d/50} - 1\right)}. \tag{3}$$

The scaled recombination rate $\rho$ was defined as in the main text as $\rho = 2Nd/100$ [4].

Consider now the no-recombination probability, $\pi_{\text{nr}}$. As long as $d \geq m$, $\pi_{\text{nr}}$ is trivially 1. If $d < m$, the segment spanning the two sites is of length $d + \ell_{1L} + \ell_{2R}$, where $\ell_{1L}$ is the distance to the next recombination event to the *left of the left marker*, and similarly for $\ell_{2R}$ (see Figure 1 for illustration). Given that the coalescence time (at both sites) was $t$, both $\ell_{1L}$ and $\ell_{2R}$ are exponentially distributed with rate $2Nt/100$. The PDF of the coalescence time is $\Phi(t) = (1 + \rho)e^{-(1+\rho)t}$, since this is the PDF of the time to exit state 1, and we are given that there was no recombination before coalescence. Therefore,

$$\pi_{\text{nr};d<m} = \int_0^{\infty} (1 + \rho)e^{-(1+\rho)t}dt \int_{m-d}^{\infty}\left(\frac{Nt}{50}\right)^2 \ell e^{-Nt\ell/50}d\ell. \tag{4}$$

These integrals are easily solvable, giving

$$\pi_{\text{nr}} = \begin{cases} 1 - \left[\frac{N(m-d)}{N(m-d)+50(1+\rho)}\right]^2 & d < m, \\ 1 & d \geq m. \end{cases} \tag{5}$$

It is easy to see that $\lim_{d\to m^+}\pi_{\text{nr}} = \lim_{d\to m^-}\pi_{\text{nr}}$, as expected.

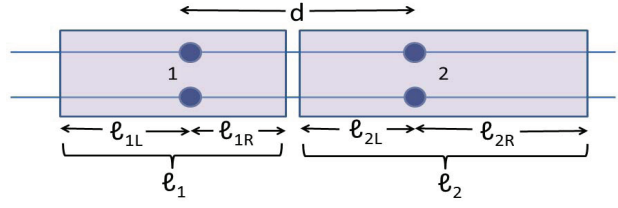

Figure 1: An illustration of the shared segments spanning two sites (numbered 1 and 2). Lines correspond to chromosomes and circles to sites, which are distance $d$ apart. The shaded boxes correspond to hypothetical shared segments. The left segment extends to distance $\ell_{1R}$ to the right of the site and $\ell_{1L}$ to the left of it, and similarly for the right segment.

The case of recombination is more complicated. One might think that if there was a recombination event in the history of the two sites, then the two sites will be shared or not independently. However, the presence of a recombination event implies that the sum of $\ell_{1R}$ and $\ell_{2L}$ [[the segment length to the right of the left marker) and (the segment length to the left of the right marker)] cannot exceed $d$ (see Figure 1 for illustration). We simplify the analysis by assuming instead that each of those two segments cannot exceed length $d$, but that their lengths are otherwise independent, resulting in a slight overestimation of $\pi_r$. Thus, for a given time to MRCA, $t_1$, the segment length spanning the left site can be written as $\ell_1 = \ell_{1L} + \ell_{1R}$ (see Figure 1), where $\ell_{1L}$ is distributed exponentially with rate $Nt_1/50$,

$$P(\ell_{1L}) = \frac{Nt_1}{50} e^{-\frac{Nt_1 \ell_{1L}}{50}} \quad ; \quad \ell_{1L} > 0, \tag{6}$$

and $\ell_{1R}$ is similarly distributed, except for an upper cutoff at $\ell_{1R} = d$,

$$P(\ell_{1R}) = \frac{\frac{Nt_1}{50} e^{-\frac{Nt_1 \ell_{1R}}{50}}}{1 - e^{-\frac{Nt_1 d}{50}}} \quad ; \quad 0 < \ell_{1R} < d. \tag{7}$$

Using convolution, the probability density function of $\ell_1 = \ell_{1L} + \ell_{1R}$ is

$$P(\ell_1) = \frac{\left(\frac{Nt_1}{50}\right)^2 e^{-\frac{Nt_1 \ell_1}{50}}}{1 - e^{-\frac{Nt_1 d}{50}}} \cdot \begin{cases} d & \ell_1 < d, \\ \ell_1 & \ell_1 \geq d. \end{cases} \tag{8}$$

The probability that $\ell_1 \geq m$ and thus the site is on a shared segment is

$$P(\ell_1 > m) = \frac{1}{1 - e^{-\frac{Nt_1 d}{50}}} \cdot \begin{cases} d\frac{Nt_1}{50} e^{-\frac{Nt_1 m}{50}} & d < m, \\ \left(1 + m\frac{Nt_1}{50}\right) e^{-\frac{Nt_1 m}{50}} - e^{-\frac{Nt_1 d}{50}} & d \geq m. \end{cases} \tag{9}$$

For large $d$, $P(\ell_1 > m) \to \left(1 + m\frac{Nt_1}{50}\right) e^{-\frac{Nt_1 m}{50}}$, which is exactly the single-site expression (Eq. (1) in the main text), as expected. We then simplify again by approximating the denominator of $P(\ell_1 > m)$ with 1,

$$P(\ell_1 > m) \approx \begin{cases} d\frac{Nt_1}{50} e^{-\frac{Nt_1 m}{50}} & d < m, \\ \left(1 + m\frac{Nt_1}{50}\right) e^{-\frac{Nt_1 m}{50}} - e^{-\frac{Nt_1 d}{50}} & d \geq m. \end{cases} \tag{10}$$

This should lead to a slight underestimation of $\pi_r$. From here on the calculation is exact. An equation identical to (10) holds for $P(\ell_2 > m)$. Integrating the probabilities of the two sites to be in shared segments over all possible coalescence times, we have, for $d < m$,

$$\pi_r = \int_0^\infty \int_0^\infty \Phi(t_1, t_2) d\frac{Nt_1}{50} e^{-\frac{Nt_1 m}{50}} d\frac{Nt_2}{50} e^{-\frac{Nt_2 m}{50}} dt_1 dt_2. \tag{11}$$

                                    S. Carmi *et al.*

For $d \geq m$,

$$\pi_{\mathrm{r}} = \int_0^\infty \int_0^\infty \Phi(t_1, t_2) \left[ \left(1 + m\frac{Nt_1}{50}\right) e^{-\frac{Nt_1 m}{50}} - e^{-\frac{Nt_1 d}{50}} \right] \left[ \left(1 + m\frac{Nt_2}{50}\right) e^{-\frac{Nt_2 m}{50}} - e^{-\frac{Nt_2 d}{50}} \right] dt_1 dt_2. \quad (12)$$

As in the main text, this can be rewritten naturally in terms of the Laplace transform of $\Phi$,

$$\widehat{\Phi}(q_1, q_2) = \int_0^\infty \int_0^\infty e^{-q_1 t_1 - q_2 t_2} \Phi(t_1, t_2) dt_1 dt_2. \quad (13)$$

After some algebra, we find, for $d < m$,

$$\pi_{\mathrm{r}} = d^2 \left[ \frac{\partial}{\partial m_1} \frac{\partial}{\partial m_2} \widehat{\Phi}\left(\frac{m_1 N}{50}, \frac{m_2 N}{50}\right) \right]_{\substack{m_1 = m \\ m_2 = m}}. \quad (14)$$

For $d \geq m$,

$$\pi_{\mathrm{r}} = \widehat{\Phi}\left(\frac{mN}{50}, \frac{mN}{50}\right) - 2m \left[ \frac{\partial}{\partial m_1} \widehat{\Phi}\left(\frac{m_1 N}{50}, \frac{mN}{50}\right) \right]_{m_1 = m} + m^2 \left[ \frac{\partial}{\partial m_1} \frac{\partial}{\partial m_2} \widehat{\Phi}\left(\frac{m_1 N}{50}, \frac{m_2 N}{50}\right) \right]_{\substack{m_1 = m \\ m_2 = m}}$$

$$+ \widehat{\Phi}\left(\frac{dN}{50}, \frac{dN}{50}\right) - 2\widehat{\Phi}\left(\frac{mN}{50}, \frac{dN}{50}\right) + 2m \left[ \frac{\partial}{\partial m_1} \widehat{\Phi}\left(\frac{m_1 N}{50}, \frac{dN}{50}\right) \right]_{m_1 = m}. \quad (15)$$

We are therefore left only with finding $\widehat{\Phi}(q_1, q_2)$. This can be carried out almost as in the main text, except that we must take into account that there was recombination before coalescence, that is, the Markov chain jumped from the initial state 1 to state 2 and not to state 8. Therefore, the coalescence times at the two sites, $t_1$ and $t_2$, can be seen as a sum of $t'$, the time it took to jump from state 1 to state 2, and the times it took from state 2 until coalescence events occurred in both sites. As we explained just before Eq. (4), the time it takes to jump from state 1 to state 2, given recombination, is distributed exponentially with rate $(1 + \rho)$. Therefore,

$$\Phi(t_1, t_2) dt_1 dt_2 = \begin{cases} \int_0^{t_1} (1 + \rho) e^{-(1+\rho)t'} P_{21}(t_1 - t') \delta(t_2 - t_1) dt' dt_1 dt_2 & t_1 = t_2, \\ \int_0^{t_1} (1 + \rho) e^{-(1+\rho)t'} \left[ P_{22}(t_1 - t') + P_{23}(t_1 - t') \right] e^{-(t_2 - t_1)} dt' dt_1 dt_2 & t_1 < t_2, \\ \int_0^{t_2} (1 + \rho) e^{-(1+\rho)t'} \left[ P_{22}(t_2 - t') + P_{23}(t_2 - t') \right] e^{-(t_1 - t_2)} dt' dt_1 dt_2 & t_2 < t_1. \end{cases} \quad (16)$$

In the last equation, $P_{2i}(t)$ is the probability of the chain to be at state $i$ at time $t$, given that it started at state 2. The reasoning behind the last equation is as follows. In the case $t_1 = t_2$, to coalesce at both sites at time $t_1$, we need to wait time $t'$ to jump to state 2, then be back in state 1 after another period of $(t_1 - t')$ (probability $P_{21}(t_1 - t')$), and then jump to state 8 (probability $dt_1$). To coalesce at site 1 (the left one) only at time $t_1$, we need to wait time $t'$ to get to state 2, and then be at state 2 (or 3) at time $(t_1 - t')$ (probability $P_{22}(t_1 - t')$ or $P_{23}(t_1 - t')$) and jump to state 5 (or 7; probability $dt_1$). Then, coalescence at site 2 (the right one) at time $t_2 > t_1$ occurs with probability $e^{-(t_2 - t_1)} dt_2$. The case $t_1 > t_2$ is similarly explained. Taking the Laplace transform of the last equation,

$$\widehat{\Phi}(q_1, q_2) = \int_0^\infty \int_0^\infty e^{-q_1 t_1 - q_2 t_2} \int_0^{t_1} (1 + \rho) e^{-(1+\rho)t'} P_{21}(t_1 - t') \delta(t_2 - t_1) dt' dt_1 dt_2 \quad (17)$$

$$+ \int_0^\infty \int_{t_1}^\infty e^{-q_1 t_1 - q_2 t_2} \int_0^{t_1} (1 + \rho) e^{-(1+\rho)t'} \left[ P_{22}(t_1 - t') + P_{23}(t_1 - t') \right] e^{-(t_2 - t_1)} dt' dt_2 dt_1$$

$$+ \int_0^\infty \int_{t_2}^\infty e^{-q_1 t_1 - q_2 t_2} \int_0^{t_2} (1 + \rho) e^{-(1+\rho)t'} \left[ P_{22}(t_2 - t') + P_{23}(t_2 - t') \right] e^{-(t_1 - t_2)} dt' dt_1 dt_2.$$

The first term of the right-hand-side can be solved as follows,

$$\int_0^\infty \int_0^\infty e^{-q_1 t_1 - q_2 t_2} \int_0^{t_1} (1+\rho) e^{-(1+\rho)t'} P_{21}(t_1 - t') dt' dt_1 \delta(t_2 - t_1) dt_2 =$$

$$(1+\rho) \int_0^\infty e^{-(q_1 + q_2)t_1} \left[ \int_0^{t_1} e^{-(1+\rho)t'} P_{21}(t_1 - t') dt' \right] dt_1 =$$

$$\frac{1+\rho}{1+\rho+q_1+q_2} \widehat{P}_{21}(q_1 + q_2). \tag{18}$$

The last line results from the special structure of the integrals in the second line: the internal integral is a convolution between $e^{-(1+\rho)t}$ and $P_{21}(t)$, and the external integral is the Laplace transform $t_1 \to (q_1 + q_2)$ of the internal integral. Applying the convolution theorem (recalling that the Laplace transform of $e^{-at}$ is $(a+q)^{-1}$), we arrive at the last line. The second and third terms of Eq. (17) require more algebra but are solved similarly, finally giving

$$\widehat{\Phi}(q_1, q_2) = \frac{1+\rho}{1+\rho+q_1+q_2} \left\{ \left( \frac{1}{1+q_1} + \frac{1}{1+q_2} \right) \left[ \widehat{P}_{22}(q_1 + q_2) + \widehat{P}_{23}(q_1 + q_2) \right] + \widehat{P}_{21}(q_1 + q_2) \right\}. \tag{19}$$

By that we are almost done, since as in the main text, the Laplace transform of the transition probabilities $\widehat{P}_{2i}(q)$ can be readily found using the continuous-time Markov chain relation

$$\widehat{P}_{2i}(q) = (qI - Q)_{2i}^{-1}, \tag{20}$$

where $Q$ is the transition rate matrix of the chain. Substituting, using MATHEMATICA, Eq. (20) in Eq. (19) gives

$$\widehat{\Phi}(q_1, q_2) = \frac{(1+\rho)\{2(6+q)[3 + q_1(4+q_1) + q_2(4+q_2) + 3q_1 q_2] + \rho(2+q)(13+3q) + \rho^2(2+q)\}}{(1+q_1)(1+q_2)(1+q+\rho)[2(1+q)(3+q)(6+q) + \rho(2+q)(13+3q) + \rho^2(2+q)]}, \tag{21}$$

where $q = q_1 + q_2$. We then substituted, again using MATHEMATICA, Eq. (21) in Eqs. (14) and (15) to obtain the final expression for $\pi_r$. We verified numerically that $\lim_{d \to m^+} \pi_r = \lim_{d \to m^-} \pi_r$. Eq. (5) for $\pi_{nr}$, Eq. (3) for $p_{nr}$, and Eq. (2) for $\pi_2$ complete the derivation.

## S1.4 An alternative derivation of $\widehat{\Phi}(q_1, q_2)$ using the Feynman-Kac formula

In this subsection, we show how $\widehat{\Phi}(q_1, q_2)$ (Eq. (11) in the main text and Eq. (21) here) can be derived using the Feynman-Kac formula as described by Fitzsimmons and Pitman [5]. We thank an anonymous reviewer for pointing out this approach.

Let us start with Eq. (11) in the main text. Assume the same continuous-time Markov chain as in the main text, and define a *functional* of the Markov chain as $A_v = \int_0^T v(X_t)dt$, where $X_t$ is the state of the chain at time $t$, $T$ is the "killing" time when the chain reaches an absorbing state (in our case, state no. 8), and $v(x)$ assigns a value to each state. With this notation, the Laplace transform $\widehat{\Phi}(q_1, q_2)$ (for the case analyzed in the main text, when there is no restriction on the first transition) can be written as

$$\widehat{\Phi}(q_1, q_2) = \int_0^\infty \int_0^\infty e^{-q_1 t_1 - q_2 t_2} \Phi(t_1, t_2) dt_1 dt_2 = \langle e^{A_v} \rangle, \tag{22}$$

with $v = -(q_1 + q_2, q_1 + q_2, q_1 + q_2, q_1, q_2, q_1, q_2)^T$. This is true, because the left-site coalescence time $t_1$ is the total time spent by the chain in states 1,2,3,4, and 6, whereas the right-site coalescence time $t_2$ is the total time spent in 1,2,3,5, and 7.

According to the Feynman-Kac formula [5],

$$\widehat{\Phi}(q_1, q_2) = \langle e^{A_v} \rangle = \lambda(Q' + M_v)^{-1}Q'\mathbf{1}, \tag{23}$$

where $M_v = \text{diag}(v)$, $\lambda$ is the initial condition (in our case, $\lambda = (1, 0, 0, 0, 0, 0, 0, 0)$, since the chain always starts at state 1), and $\mathbf{1} = (1, 1, 1, 1, 1, 1, 1, 1)^T$. The matrix $Q'$ is obtained from the transition rate matrix $Q$ by removing the row and column corresponding to the absorbing state (state 8). Carrying out the necessary matrix multiplications and inversions, we obtain the exact same expression as in Eq. (11) in the main text.

In the case analyzed in Section S1.3 above (leading eventually to Eq. (21)), the chain is guaranteed to jump from state 1 to state 2 (but not to state 8) at rate $(1 + \rho)$. This can be incorporated into the Feynman-Kac framework by extending the chain to include a "ghost" state 0, from which the only outward transition is to state 2, at rate $(1 + \rho)$. No transitions are allowed into state 0, and it is the initial state of the chain. Since neither site has coalesced while in state 0, we can write $\widehat{\Phi}(q_1, q_2) = \langle e^{A_v} \rangle$ with $v = -(q_1 + q_2, q_1 + q_2, q_1 + q_2, q_1 + q_2, q_1, q_2, q_2)^T$. We then use $\langle e^{A_v} \rangle = \lambda (Q'' + M_v)^{-1} Q'' \mathbf{1}$, where $\lambda = (1, 0, 0, 0, 0, 0, 0, 0)$ and $Q''$ is equal to $Q'$, but with an additional row and an additional column for the new state 0:

$$Q'' = \begin{pmatrix} -1-\rho & 0 & 1+\rho & 0 & 0 & 0 & 0 & 0 \\ 0 & & & & & & & \\ 0 & & & & & & & \\ 0 & & & & & & & \\ 0 & & & & Q' & & & \\ 0 & & & & & & & \\ 0 & & & & & & & \\ 0 & & & & & & & \end{pmatrix}. \tag{24}$$

Solving and simplifying gives Eq. (21).

## S1.5   A linearly expanding population

In this subsection we calculate the mean and the variance of the total sharing for a linearly expanding population. Define the population size as $N(t) = N_0 \lambda(t)$, where

$$\lambda(t) = \begin{cases} 1 + \tilde{r}(t_0 - t) & 0 \le t \le t_0, \\ 1 & t > t_0. \end{cases} \tag{25}$$

This corresponds to a population maintaining a constant size until $t = t_0$ generations ago; starting at $t = t_0$ and until present, the population grows linearly at rate $\tilde{r}$. The PDF of the coalescence times is

$$\Phi(t) = \frac{e^{-\int_0^t \frac{dt'}{\lambda(t')}}}{\lambda(t)}. \tag{26}$$

Substituting $\lambda(t)$ from Eq. (25), we have, for $t \le t_0$,

$$\begin{aligned} \Phi(t) &= \frac{1}{1 + \tilde{r}(t_0 - t)} \exp\left[ -\int_0^t \frac{dt'}{1 + \tilde{r}(t_0 - t')} \right] \\ &= \frac{1}{1 + \tilde{r}(t_0 - t)} \exp\left\{ \frac{1}{\tilde{r}} \ln\left[ \frac{1 + \tilde{r}(t_0 - t)}{1 + \tilde{r}t_0} \right] \right\} \\ &= (1 + \tilde{r}t_0)^{-1/\tilde{r}} [1 + \tilde{r}(t_0 - t)]^{1/\tilde{r} - 1}. \end{aligned} \tag{27}$$

For $t > t_0$,

$$\begin{aligned} \Phi(t) &= \exp\left[ -\int_0^{t_0} \frac{dt'}{1 + \tilde{r}(t_0 - t')} - \int_{t_0}^t dt' \right] \\ &= \exp\left[ \frac{1}{\tilde{r}} \ln\left( \frac{1}{1 + \tilde{r}t_0} \right) - (t - t_0) \right] \\ &= (1 + \tilde{r}t_0)^{-1/\tilde{r}} e^{-(t - t_0)}. \end{aligned} \tag{28}$$

In summary,

$$\Phi(t) = (1 + \tilde{r}t_0)^{-1/\tilde{r}} \begin{cases} [1 + \tilde{r}(t_0 - t)]^{1/\tilde{r}-1} & 0 \le t \le t_0 \\ e^{-(t-t_0)} & t > t_0 \end{cases}. \tag{29}$$

We then use Eq. (17) from the main text for the mean total sharing,

$$\langle f_T \rangle = \int_0^\infty \Phi(t) \left(1 + \frac{mN_0 t}{50}\right) e^{-\frac{mN_0 t}{50}} dt, \tag{30}$$

and Eq. (19) from the main text for the variance of the total sharing,

$$\mathrm{Var}\,[f_T] \approx 2 \int_{m/L}^1 (1 - x) \left[\int_0^\infty \Phi(t) e^{-txN_0 L/50} dt\right] dx. \tag{31}$$

The integral in $\langle f_T \rangle$ and the internal integral (over $t$) in $\mathrm{Var}\,[f_T]$ can be evaluated in terms of incomplete Gamma functions (not shown). For $\mathrm{Var}\,[f_T]$, the external integral must be evaluated numerically. [We also tried to change the order of the integration in Eq. (31), that is, to compute the integral over $x$ first. However, in that case, while the integral over $x$ was solvable, the integral over $t$ was not.] We compare the results of Eqs. (30) and (31) to simulations in Figure **S1**. In the simulations, the ancestral population size was set to $N_a = 10000$, the expansion started $E_t = 500$ generations ago, and the final (current) population size varied in the range $N_c = [10500, 15000]$. In terms of the parameters of $\lambda(t)$, this corresponds to $N_0 = N_a = 10000$, $t_0 = 500/10000 = 0.05$, and $\tilde{r}$ between $(1.05 - 1)/0.05 = 1$ and $(1.5 - 1)/0.05 = 10$. The comparison shows reasonable agreement with deviation of up to about 10%.

## S2 The distribution of the total sharing

This section provides some additional results and discussion on *The total sharing distribution and an error model* section in the main text, in which an approximation to the distribution of the total sharing was presented.

### S2.1 A bound on the probability of no sharing

A bound on the probability of no sharing, $P(f_T = 0)$, can be obtained directly from the one-sided Chebyshev inequality,

$$P\left(f_T \le \langle f_T \rangle - a\right) \le \frac{\sigma_{f_T}^2}{\sigma_{f_T}^2 + a^2}. \tag{32}$$

Substituting $a = \langle f_T \rangle$ and noting that $P(f_T \le 0) = P(f_T = 0)$ immediately gives

$$P(f_T = 0) \le \frac{\sigma_{f_T}^2}{\sigma_{f_T}^2 + \langle f_T \rangle^2}. \tag{33}$$

In practice, however, this bound is not very tight, as can be seen in Figure **S3**.

### S2.2 IBD calculations in the founder model

*The total sharing distribution and an error model* section in the main text presented results for the distribution of total sharing assuming it is a sum of a Poisson distributed number of segments. Early calculations of the distribution of the total sharing were performed in a different population model, where a group of unrelated individuals is assumed to have recently founded the population. The distribution of the total length of the IBD shared segments was calculated, under somewhat strong assumptions, using renewal theory [6, 7]. In their model, it was assumed that if a region is not shared IBD, it is fully heterozygous (because it is

 

derived from different founders). In reality, however, all segments descend from a common ancestor at some point in the past, but the common ancestor of some segments is so ancient that they are too short to be detected. Our coalescent-based approach takes just that into account, by considering as IBD only segments longer than a certain length threshold.

### S2.3  Matching the Poisson and exponential parameters

The parameters of the Poisson approximation, Eq. (24) in the main text, can be obtained by matching the first two moments of the total sharing distribution. The mean and variance of the Poisson approximation are given by (see, e.g., the main text Eq. (25))

$$
\begin{aligned}
\langle L_T \rangle &= n_0(\ell_0 + m) = L \langle f_T \rangle, \\
\mathrm{Var}\left[L_T\right] &= n_0[\ell_0^2 + (\ell_0 + m)^2] = L^2 \sigma_{f_T}^2,
\end{aligned}
\tag{34}
$$

where $\langle f_T \rangle$ is given in the main text Eq. (4) and $\sigma_{f_T}$ is given by one of the previously calculated approximations, e.g., the main text Eq. (15). Solving for $n_0$ and $\ell_0$ in terms of $\langle f_T \rangle$ and $\sigma_{f_T}$ gives

$$
\begin{aligned}
\ell_0 &= \frac{L\sigma_{f_T}^2 - 2\langle f_T \rangle m + \alpha}{4\langle f_T \rangle}, \\
n_0 &= \frac{L\sigma_{f_T}^2 + 2\langle f_T \rangle m - \alpha}{2m^2/L},
\end{aligned}
\tag{35}
$$

where $\alpha = \left(4\langle f_T \rangle Lm\sigma_{f_T}^2 + L^2\sigma_{f_T}^4 - 4\langle f_T \rangle^2 m^2\right)^{1/2}$. In practice, we found that using Eq. (35) matched well the distribution $P(f_T)$ only when we underestimated $\sigma_{f_T}$ by 20-30%, probably because of the absence of the broad tail in Eq. (24). Therefore, in Figures 4 in the main text and **S2** here we used the fitted values of $n_0$ and $\ell_0$.

## S3  An estimator of the population size

In this subsection, we derive Eq. (39) in the main text for the variance of an estimator of the population size that is based on the average sharing between all pairs in a cohort. For a cohort of size $n$, define $\overline{\overline{f_T}} = \sum_{i=1}^{n}\sum_{j>i}^{n} f_T^{(i,j)}/\binom{n}{2}$, or

$$
\overline{\overline{f_T}} = \frac{f_T^{(1,2)} + f_T^{(1,3)} + \cdots f_T^{(1,n)} + f_T^{(2,3)} + \cdots + f_T^{(2,n)} + \cdots + f_T^{(n-1,n)}}{\binom{n}{2}}.
\tag{36}
$$

The estimator takes the form

$$
\hat{N} = \frac{100}{m\overline{\overline{f_T}}} - \frac{75}{m}.
\tag{37}
$$

The SD of $\hat{N}$ can be approximated as in the main text,

$$
\sigma_{\hat{N}} \approx \frac{100}{m} \frac{\sigma_{\overline{\overline{f_T}}}}{\left\langle \overline{\overline{f_T}} \right\rangle^2}.
\tag{38}
$$

In fact, this approximation is better justified here than in the main text, as the distribution of $\overline{\overline{f_T}}$ is much narrower than that of $f_T$. Using $\left\langle \overline{\overline{f_T}} \right\rangle = \langle f_T \rangle \approx 100/(mN)$ gives

$$
\sigma_{\hat{N}} \approx \frac{mN^2\sigma_{\overline{\overline{f_T}}}}{100}.
\tag{39}
$$

We therefore need to calculate the variance of $\overline{\overline{f_T}}$, from which we will then obtain the standard deviation $\sigma_{\overline{\overline{f_T}}}$. The variance of $\overline{\overline{f_T}}$ can be written as

$$\text{Var}\left[\overline{\overline{f_T}}\right] = \text{var term} + \text{cov term},\tag{40}$$

where the var term corresponds to the variances of the individual terms in the sum in the definition of $\overline{\overline{f_T}}$ (Eq. (36)), and the cov term corresponds to the covariances of these terms. More concretely, using Eq. (36),

$$\text{var term} = \frac{\binom{n}{2}\sigma_{f_T}^2}{\binom{n}{2}^2} = \frac{\sigma_{f_T}^2}{\binom{n}{2}} \approx \frac{2\sigma_{f_T}^2}{n^2} \approx \frac{2\cdot 100}{n^2 NL}\ln\left(\frac{L}{m}\right),\tag{41}$$

where we used Eq. (15) in the main text for $\sigma_{f_T}^2$. The covariance term is

$$\text{cov term} = \frac{\sum_{(i,j),i\neq j}\sum_{(k,l)\neq(i,j),k\neq l}\text{Cov}\left[f_T^{(i,j)},f_T^{(k,l)}\right]}{\binom{n}{2}^2}.\tag{42}$$

Note that the set $(i,j,k,l)$ must have at least three distinct indexes. In most combinations of $(i,j,k,l)$, we will have all $i,j,k,l$ different, for which we assume that the covariance $\text{Cov}\left[f_T^{(i,j)},f_T^{(k,l)}\right]$ is zero. We therefore have to consider only covariances of the form $\text{Cov}\left[f_T^{(i,j)},f_T^{(i,k)}\right]$ and $\text{Cov}\left[f_T^{(i,j)},f_T^{(j,k)}\right]$. Since for each pair $(i,j)$ (from which we have $\binom{n}{2}$) there are $(n-2)$ possible $k$s, we have

$$\text{cov term} \approx \frac{\binom{n}{2}2(n-2)\text{Cov}\left[f_T^{(1,2)},f_T^{(1,3)}\right]}{\binom{n}{2}^2} \approx \frac{4\text{Cov}\left[f_T^{(i,j)},f_T^{(i,k)}\right]}{n} \approx \frac{4\cdot 10000}{nN^2 mL},\tag{43}$$

where we used Eq. (27) in the main text for $\text{Cov}\left[f_T^{(1,2)},f_T^{(1,3)}\right]$. In total, the variance of $\overline{\overline{f_T}}$ is

$$\text{Var}\left[\overline{\overline{f_T}}\right] \approx \frac{2\cdot 100}{n^2 NL}\ln\left(\frac{L}{m}\right) + \frac{4\cdot 10000}{nN^2 mL} = \frac{400}{nNL}\left[\frac{\ln\left(\frac{L}{m}\right)}{2n} + \frac{100}{Nm}\right],\tag{44}$$

and

$$\sigma_{\overline{\overline{f_T}}} \approx \frac{20}{\sqrt{nNL}}\sqrt{\frac{\ln\left(\frac{L}{m}\right)}{2n} + \frac{100}{Nm}}.\tag{45}$$

Finally,

$$\sigma_{\hat{N}} \approx \frac{mN^2\sigma_{\overline{\overline{f_T}}}}{100} \approx \frac{mN^{3/2}}{5\sqrt{nL}}\sqrt{\frac{\ln\left(\frac{L}{m}\right)}{2n} + \frac{100}{Nm}},\tag{46}$$

which is precisely Eq. (39) in the main text.

## S4 An admixture pulse

In the main text, an approximate solution was given for the integral in Eq. (43). The full solution is:

$$\text{Var}\left[f_T\right] \approx 2\int_{m/L}^1 (1-x)\left[\int_0^{T_a} e^{-t-txNL/50}dt\right]dx + 2\alpha^2\int_{m/L}^1 (1-x)\left[\int_{T_a}^\infty e^{-t-txNL/50}dt\right]dx$$
$$= \frac{100}{L^2 N^2 T_a}\left\{50(1-\alpha^2)\left[\exp\left(-\frac{T_a(50+NL)}{50}\right) - \exp\left(-\frac{T_a(50+Nm)}{50}\right)\right] - NT_a(L-m)\right.$$
$$\left. + T_a(50+NL)\ln\left(\frac{50+NL}{50+Nm}\right) + T_a(50+NL)(1-\alpha^2)\left[E_i\left(-\frac{T_a(50+Nm)}{50}\right) - E_i\left(-\frac{T_a(50+NL)}{50}\right)\right]\right\},\tag{47}$$

where $E_i(x)$ is the exponential integral function. To obtain the simplified equation (43) of the main text, we assumed $T_a \ll 1$ (or $G_a = NT_a \ll N$), $m \ll L$, $mNT_a \ll 50$, $mN \gg 50$, and $LNT_a \gg 50$, and used the series expansion of the exponential integral. For the parameters for which we plotted the simulation results, the simplified expression deviates in no more than 1% from the full expression.

Simulations for the case of pulse admixture were performed using GENOME as described in the main text, with the following population history. The initial (current) population size was set to $N$, followed by splitting to two populations, at generation $G_a$, of relative sizes $N\alpha/(1 - \alpha)$ and $N$, such that a fraction $\alpha$ of the lineages descends from the first population (we could not find a way to implement the gene flow in GENOME while keeping the population size fixed). At the next generation, the first population size was reduced back to $N$, and the second was increased to $10^6$, to practically eliminate IBD sharing within the second population. At generation $10^4$, the two populations were merged into a single population of size $N$, to enable all lineages to coalesce. Simulation results are presented in Figure **S10**A. Each data point corresponds to 500 runs. The apparent noise for large $\alpha$ might be due to this somewhat unnatural admixture model implementation.

**File S2**

Supplementary Code

Matlab code for the main results

 S. Carmi *et al.*

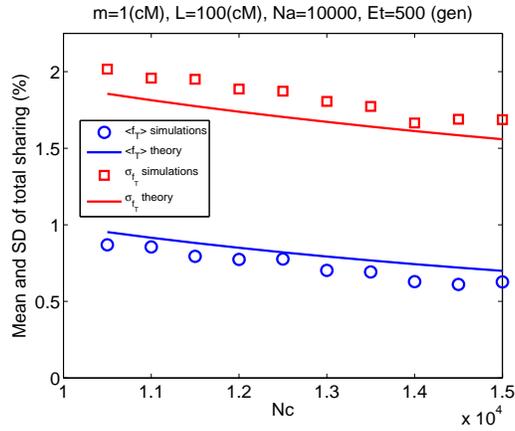

**Figure S1**: Simulation results for a linearly expanding population. Simulation results (symbols) are shown for the mean and the standard deviation (SD) of the total sharing vs. the current population size $N_c$ for a linearly expanding population (ancestral population size $N_a = 10000$ until time $E_t = 500$ generations ago, then a linear expansion until the indicated current size). The theoretical curves are taken from Eq. S30 for the mean and Eq. S31 for the SD, along with Eq. S29 for the coalescence time PDF, $\Phi(t)$. The integrals were evaluated (analytically wherever possible; see File S1) in MATHEMATICA.

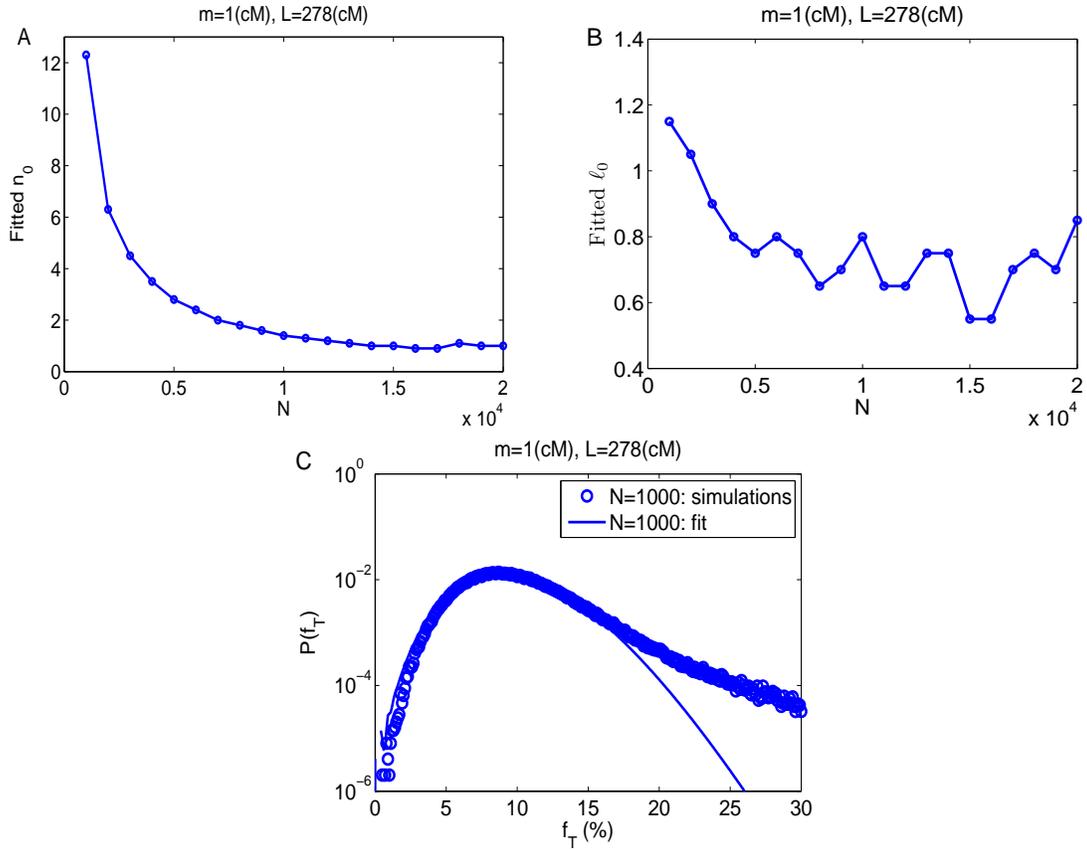

**Figure S2**: Fitting the distribution of the total sharing. (A) and (B) The fitted values of the compound Poisson parameters: $n_0$ (A), the average number of segments, and $\ell_0$ (B), the parameter of the shifted exponential distribution of the segment lengths ($\ell_0 + m$ is the average segment length). The parameters, which appear in the approximate distribution of the total sharing, Eq. (24) in the main text, are plotted vs. $N$. Data correspond to Figures 4A and B in the main text. The figure shows that $n_0$ roughly decreases as $1/N$, while $\ell_0$ decreases for small $N$ but then approaches a constant. (C) Same as Figure 4A in the main text, but magnified and plotted in log-scale. The fitted line, corresponding to the compound Poisson (Eq. (24) in the main text), provides a good fit to the central part of the curve, but it predicts a right tail much narrower than actually observed.

                        S. Carmi *et al.*

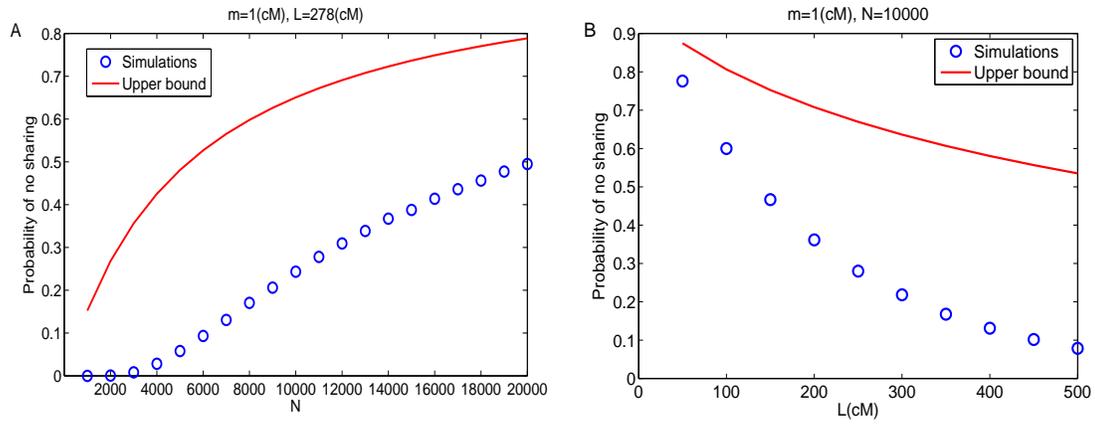

**Figure S3**: The upper bound on the probability of no sharing. Simulation results (symbols) are plotted for the fraction of pairs in the Wright-Fisher population that did not share even a single segment of length $\geq m$. Lines correspond to the theoretical upper bound, Eq. S33. (A) The probability of no sharing vs. the population size $N$ (cf. Figure 2A in the main text). (B) The probability of no sharing vs. the chromosome size $L$ (Figure 2C in the main text).

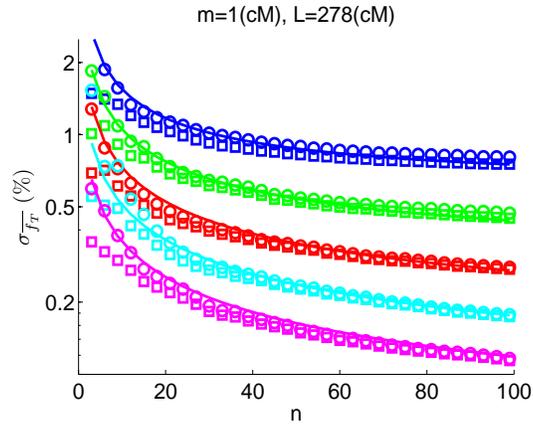

**Figure S4**: The standard deviation (SD) of the cohort-averaged sharing. Simulation results for $\sigma_{\overline{f_T}}$, the SD of the cohort-averaged sharing (in percentage of the genome) vs. the cohort size $n$. The different curves correspond to (top to bottom): $N = 1000, 2000, 4000, 8000, 16000$. Lines correspond to Eq. (28) of the main text. Squares: the SD of the cohort-averaged sharing within each cohort of $n = 100$ individuals, averaged over 100 realizations of the simulations. Circles: for comparison, the data of Figure 6A of the main text, where the cohort-averaged sharing from all realizations and all individuals was first pooled, and only then the SD was calculated. For small $n$, the average over all realizations gives a smaller variance than when pooling, but is otherwise in agreement with the prediction. The agreement is likely because as long as $n$ is not too small, the ancestral processes seen by different individuals in the cohort are only weakly correlated, and therefore the variance as calculated in the main text (over all ancestral processes) gives the correct result.

                    S. Carmi *et al.*

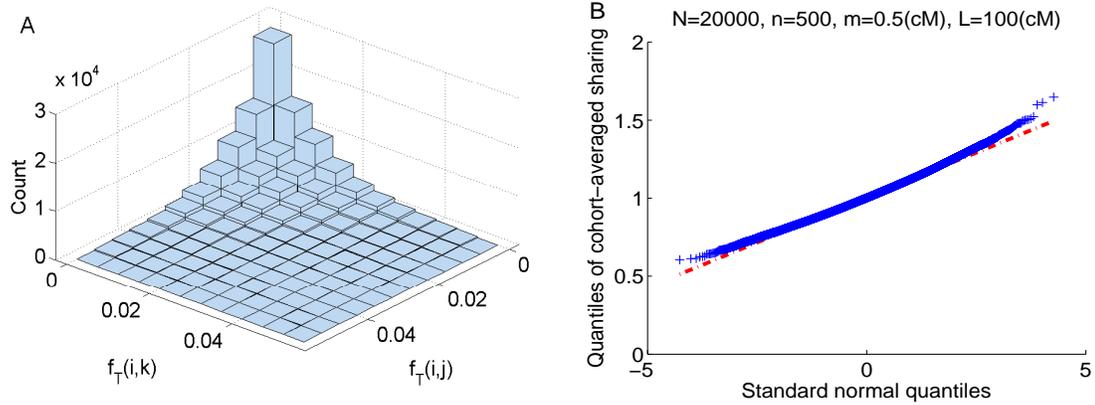

**Figure S5**: The distribution of the cohort-averaged sharing. (A) The joint distribution of the 3-way total sharing $P\left(f_T^{(i,j)}, f_T^{(i,k)}\right)$. To investigate whether the sharing fractions between two individuals to a third one depend on each other, we simulated the total sharing in populations with $N = 10000$, $m = 1$cM, $n = 100$, and one chromosome of length $L = 278$cM. For each population, we recorded all distinct values of $f_T^{(i,j)}$ and $f_T^{(i,k)}$ and plotted their joint histogram (after binning). The dependence is weak, but cannot be rejected based on a $\chi^2$-test of independence (P-value 0.12). (B) A QQ-plot of the distribution of the cohort-averaged sharing. Simulation results correspond to Figure 6B in the main text. Briefly, we calculated the distribution of the cohort-averaged sharing for populations with $N = 20000$ individuals and one chromosome of size $L = 100$cM. The minimal segment length was $m = 0.5$cM and the cohort size was $n = 500$. A QQ-plot of the distribution is shown, comparing the empirical distribution to a normal one. The distribution is quite close to normal in the central part, but with a broader right tail and a narrower left tail than expected.

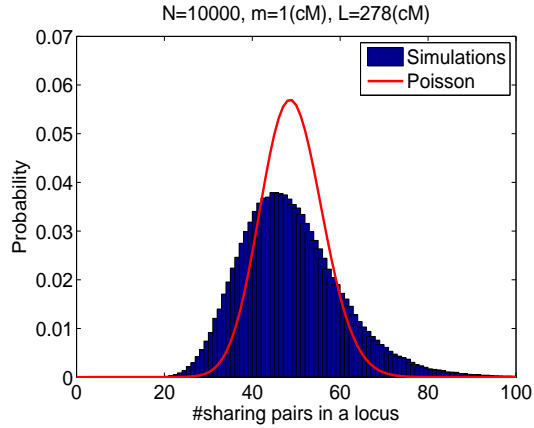

**Figure S6**: A histogram of the number of pairs sharing at each locus. We simulated 100 Wright-Fisher populations with $N = 10000$, $n = 100$, and one chromosome of length $L = 278$cM, and searched for IBD shared segments using $m = 1$cM. In the Genome coalescent simulator, recombination is resolved only within blocks whose size we set to 0.01cM. For each such block (excluding the first and last $m$(cM) of the chromosome), we recorded the number of pairs sharing a segment containing it, and then plotted the histogram over all blocks. We also plot a Poisson PDF with the same mean as the observed distribution. The histogram is significantly broader than the Poisson (Index-of-Dispersion test P-value less than Matlab's resolution), indicating that sharing tends to concentrate at specific loci.

 S. Carmi *et al.*

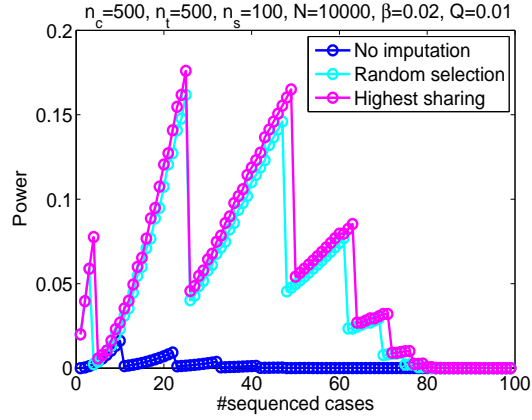

**Figure S7**: Power to detect an association when imputing by IBD. We plot the power to detect an association of a variant that exists in cases only, with and without imputation by IBD, and with sequenced individuals selected either randomly or according to their total sharing. This corresponds to the model of *Implications to sequencing study design* section in the main text. The parameters we used were: $N = 10000$, $L = 278$cM (one chromosome), $m = 1$cM, cohort size of $n_c = 500$ cases and $n_t = 500$ controls, and a total sequencing budget of $n_s = 100$ individuals. The carrier frequency here is $\beta = 0.02$, and the threshold P-value is $Q = 0.01$. For each number of sequenced cases (x-axis), $n_{c,s}$ (where due to the budget limit, the number of sequenced controls is $n_{t,s} = n_s - n_{c,s}$), we plot the power according to Eqs. (32), (34), (35), and (36) in the main text. The power vs. $n_{c,s}$ has a sawtooth shape. This was also documented in [8], where the same model was analyzed. The sawtooth is an effect of the discreteness of the model. For several different values of $n_{c,s}$ (or $n_c^{(\text{eff})}$, for that matter), the minimal number of carriers $b^*$ required to reject the null hypothesis is the same, but the probability to observe that number of carriers increases with $n_{c,s}$. As $n_{c,s}$ increases further, $b^*$ finally increases by one, reducing the power dramatically.

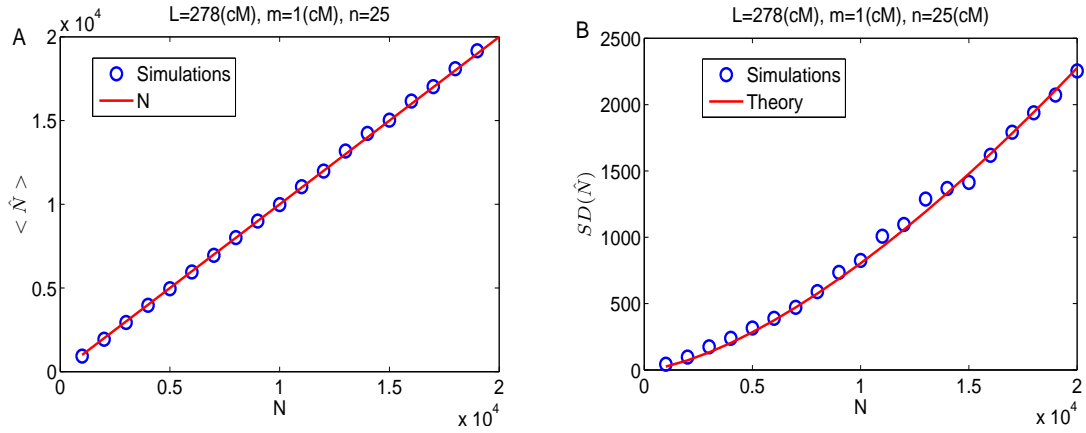

**Figure S8**: The mean and the variance of an estimator of the population size vs. $N$. We plot simulation results (symbols) for the estimator of the population size, $\hat{N}$, given in Eq. (38) in the main text. For each value of $N$, we simulated a number of Wright-Fisher populations and calculated the total sharing as in Figure 2A in the main text. For each of the populations simulated for each $n$, we divided the individuals into four disjoint groups of 25 individuals each. In each group, we calculated the mean total sharing, $\overline{\overline{f}}$, between all $\binom{25}{2}$ pairs. We then applied the main text Eq. (38) to calculate the population size estimator $\hat{N}$. Finally, for each $N$, we plotted the average of the estimator over all groups, $\left\langle \hat{N} \right\rangle$ (A), as well as its standard deviation (B). In (A), we also plot the identity line ($\left\langle \hat{N} \right\rangle = N$), and in (B), we also plot the theory, the main text Eq. (39).

                                S. Carmi *et al.*

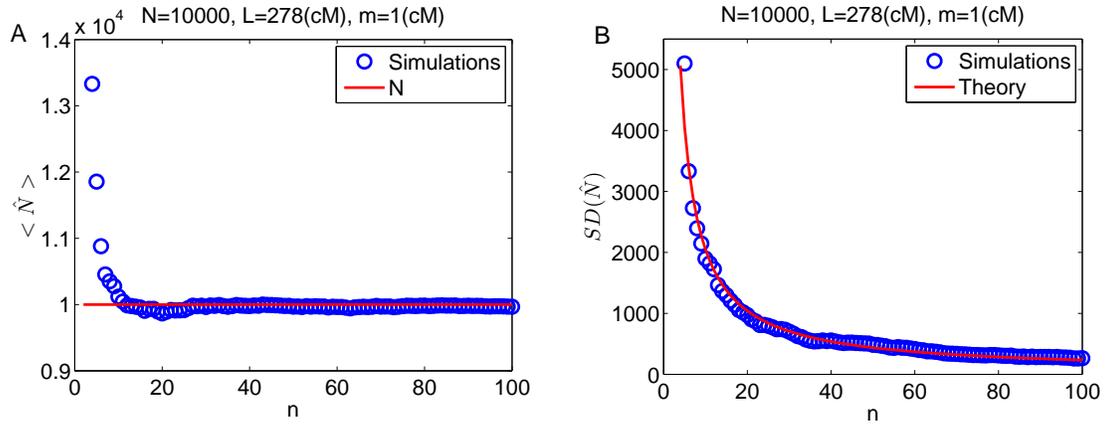

**Figure S9**: The mean and the variance of an estimator of the population size vs. $n$. This figure is as Figure **S8**, except that here the mean (A) and standard deviation (B) of the estimator $\hat{N}$ are plotted vs. the number of individuals $n$ and the population size is fixed, $N = 10000$. For each $n$, the total sharing between all pairs in a subset of $n$ individuals from each population was averaged to obtain $\overline{\overline{f}}$, and then $\hat{N}$ was calculated according to Eq. (38) in the main text. Panel (A) also shows a horizontal line at $\left\langle \hat{N} \right\rangle = N$, demonstrating that $N$ is overestimated, but only for small $n$. Panel (B) also shows the theoretical standard deviation from the main text Eq. (39).

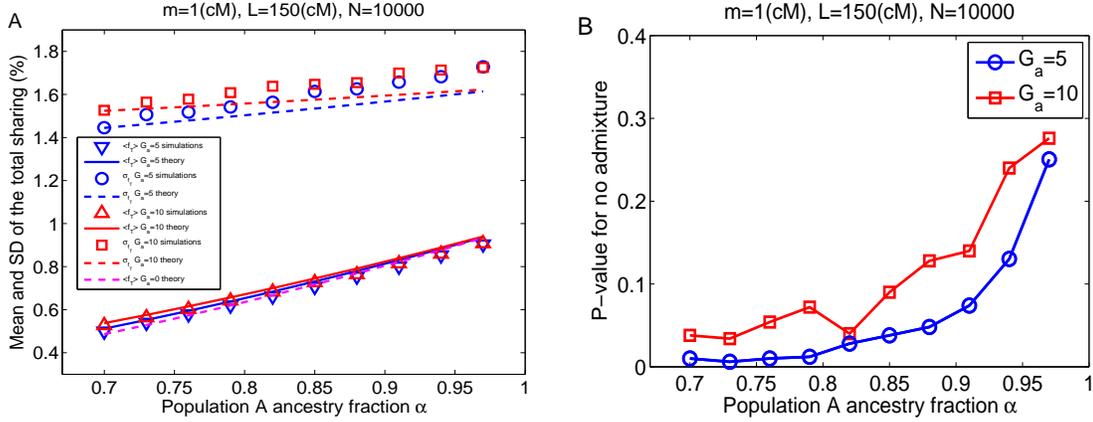

**Figure S10**: IBD sharing in an admixture pulse model. (A) Simulations were carried out for the admixture pulse model as described in the Supplementary text (File S1). The mean and the standard deviation (SD) of the total fraction of IBD sharing are plotted. Symbols correspond to simulations (triangles: mean, circles/squares: SD). Solid lines correspond to the main text Eq. (42) for the mean sharing and dashed lines to the main text Eq. (43) for the SD. Blue and red symbols/lines correspond to $G_a = 5$ and $G_a = 10$, respectively. The magenta dashed line corresponds to the theoretical mean sharing if admixture has just occurred, $G_a = 0$. (B) P-values for the admixture test. We simulated the admixture pulse model with population size of $N = 10000$, $L = 150$cM, $m = 1$cM, $G_a$ equals 5 or 10, and various values of $\alpha$. For each $G_a$ and $\alpha$, the (true) IBD shared segments were extracted and the population size was inferred as $\hat{N} = 100/(m\overline{\overline{f_T}}) - 75/m$, where $\overline{\overline{f_T}}$ is the average fraction of sharing over all pairs. Then, 500 populations were simulated with constant size $\hat{N}$, and the SD of the cohort-averaged sharing was calculated. The P-value is the fraction of times the SD in the simulations was higher than the one in the admixed population.

 S. Carmi *et al.*

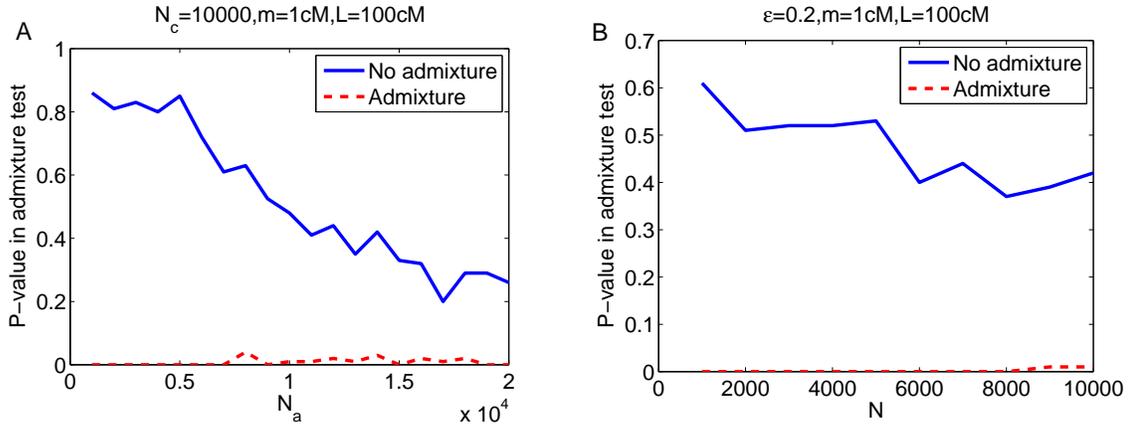

**Figure S11**: The effect of possible confounders on the admixture test. (A) The effect of a variable population size. We simulated a simple two-size population history, with ancestral population size $N_a$ until $T = 50$ generations ago, followed by population size of $N_c = 10000$ until present. $N_a$ varied between 1000 and 20000, such that both expansions and contractions were studied. We simulated two scenarios: one without admixture and one with admixture taking place $G_a = 5$ generations ago, replacing a proportion $1 - \alpha = 0.3$ of the population. We than ran the admixture test as described in the main text and in Figure **S10** (with 100 simulated constant-size populations). The results demonstrate that for all values of $N_a$ tested, while for the no-admixture case, the test always resulted in an insignificant P-value, for the admixture case, the P-value was always below 0.05. We note, however, that it might be that a more extreme or complex demographic history will confound the admixture test; but at least for the parameters investigated here, the admixture test is robust. (B) The effect of IBD detection errors. We simulated populations of constant size $N$ and dropped each detected IBD segment with probability $\epsilon = 0.2$ (as in the error model of *The total sharing distribution and an error model* section of the main text). Again, we simulated two scenarios: with and without admixture (same parameters as in (A)). We then ran the admixture test, and as in (A), the resulting P-values were significant only for the truly admixed populations.